\documentclass[twocolumn]{aastex631}

\usepackage{xspace}

\newcommand {\nustar}{\textit{NuSTAR}\xspace}
\newcommand {\nicer}{{NICER}\xspace}
\newcommand {\chandra}{\textit{Chandra}\xspace}

\newcommand {\ixpe}{\textit{IXPE}\xspace}
\newcommand {\swift}{\textit{Swift}--XRT\xspace}
\newcommand {\xmm}{\textit{XMM--Newton}\xspace}
\newcommand {\maxi}{\text{MAXI}\xspace}
\newcommand {\source}{{GX~13$+$1}\xspace}
\shorttitle{X-ray Dips and Polarization Angle Swings in \source}
\shortauthors{Di Marco et al.}
\begin{document}

\title{X-ray Dips and Polarization Angle Swings in \source}

\correspondingauthor{Alessandro Di Marco}
\email{alessandro.dimarco@inaf.it}
\author[0000-0003-0331-3259]{Alessandro Di Marco}
\affiliation{INAF Istituto di Astrofisica e Planetologia Spaziali, Via del Fosso del Cavaliere 100, 00133 Roma, Italy}
\author[0000-0001-8916-4156]{Fabio La Monaca}
\affiliation{INAF Istituto di Astrofisica e Planetologia Spaziali, Via del Fosso del Cavaliere 100, 00133 Roma, Italy}
\affiliation{Dipartimento di Fisica, Universit\`{a} degli Studi di Roma ``Tor Vergata'', Via della Ricerca Scientifica 1, 00133 Rome, Italy}
\author[0009-0009-3183-9742]{Anna Bobrikova}
\affiliation{Department of Physics and Astronomy, 20014 University of Turku,  Finland}
\author[0000-0002-0018-1687]{Luigi Stella}
\affiliation{INAF Osservatorio Astronomico di Roma, Via Frascati 33, 00078 Monte Porzio Catone (RM), Italy}
\author[0000-0001-6289-7413]{Alessandro Papitto}
\affiliation{INAF Osservatorio Astronomico di Roma, Via Frascati 33, 00078 Monte Porzio Catone (RM), Italy}
\author[0000-0002-0983-0049]{Juri Poutanen}
\affiliation{Department of Physics and Astronomy, 20014 University of Turku,  Finland}
\author[0000-0003-1285-4057]{Maria Cristina Baglio}
\affiliation{INAF--Osservatorio Astronomico di Brera, Via Bianchi 46, I-23807 Merate (LC), Italy}
\author[0000-0002-4576-9337]{Matteo Bachetti}
\affiliation{INAF Osservatorio Astronomico di Cagliari, Via della Scienza 5, 09047 Selargius (CA), Italy}
\author[0000-0001-6894-871X]{Vladislav Loktev}
\affiliation{Department of Physics and Astronomy, 20014 University of Turku,  Finland}
\author[0000-0001-7397-8091]{Maura Pilia}
\affiliation{INAF Osservatorio Astronomico di Cagliari, Via della Scienza 5, 09047 Selargius (CA), Italy}
\author[0000-0002-5359-9497]{Daniele Rogantini}
\affiliation{Department of Astronomy and Astrophysics, University of Chicago, Chicago, IL 60637, USA}



\begin{abstract}

We present the result from the April 2024 observation of the low-mass X-ray binary \source with the Imaging X-ray Polarimetry Explorer (\ixpe), together with \nicer and \swift coordinated observations. Two light curve dips were observed; during them, the harder Comptonized spectral component was dominant and the polarization degree higher than in the softer, off-dip intervals.
Through a joint analysis of the three \ixpe observations, which also included the dip from the first observation, we demonstrate that the polarization properties varied in response to the intensity and spectral hardness changes associated with the dips. The polarization degree attained values up to ${\sim}$4\%. The polarization angle showed a swing of $\sim 70^\circ$ across the dip and off-dip states, comparable to the continuous rotation seen during the first \ixpe observation.  
We discuss these results in the context of models for polarized emission from the accretion disk and the boundary/spreading layer on the neutron star surface.
We also draw attention to the role that an extended accretion disk corona or disk wind can play in generating high polarization degrees and, possibly, swings of the polarization angle. 
\end{abstract}

\keywords{accretion, accretion disks -- polarization --  stars: individual: GX~13+1 -- stars: neutron -- X-ray binaries}


\section{Introduction} \label{sec:intro}

Low Mass X-ray Binaries (LMXBs) are systems in which a compact object accretes matter from a low mass companion star via Roche-lobe overflow and a disk. Systems hosting a weakly magnetized neutron star (WMNS) comprise two main classes, named Z and Atoll sources, after the characteristic shape they trace in the X-ray color-color diagram (CCD). 
Z sources are persistent and bright, displaying luminosities close to the Eddington limit, whereas Atoll sources are about a decade less luminous, frequently display thermonuclear bursting activity and in some cases undergo transient outbursts; see, {e.g.}, \cite{vanderklis89}.
The millisecond spin periods that are expected to characterize WMNSs in LMXBs have been detected only from a few dozen Atoll sources through coherent pulsations during transient outbursts and/or burst oscillations during thermonuclear bursts \citep{Papitto2014}. Fast aperiodic and quasi-periodic variability is detected in the range between few Hz and ${\sim} 1$\,kHz and presents clear similarities in both Z and Atoll sources. 
The energy spectra of accreting WMNSs can be described by a model consisting of a multicolor blackbody, representing the thermal emissions from the inner accretion disk regions, and a higher-energy Comptonized continuum, which dominates the spectrum and likely originates from the boundary layer between the disk and the NS \citep[BL;][]{Shakura88,Popham01} or the plasma layer at the NS surface, the so-called spreading layer \citep[SL;][]{inogamov1999,suleimanov2006}. Reflection spectrum \citep{White88,George1991, Fabian1989, Giridharan2023,Ludlam2024} caused by radiation intercepted and reprocessed by the accretion disk is also observed in many cases. The presence of powerful outflows, in the form of collimated fast jets emitted perpendicularly to the disk \citep{Baglio17} and/or fan-shaped slower winds launched from moderate altitudes above the disk \citep{Allen2018} have also been established in a number of BH and NS LMXBs.

X-ray flux modulations at the orbital period are sometimes present, especially in higher inclination (${>}60\degr$) systems. In most cases, the modulation consists of intensity dips arising from obscuration due to increased photoelectric absorption, repeated once (or twice) per orbit, which are variable in shape and present some jitter in the orbital phase. These dips are likely due to clumps in the outermost disk region, close to the point where the accretion stream from the companion impacts. X-ray eclipses resulting from direct occultation of the emitting regions by the companion star are rare phenomena. Residual X-ray flux is observed at the bottom of eclipses ({e.g.}, in \mbox{EXO 0748$-$676}). This, together with the smooth, often low-amplitude, orbital modulation observed in some systems, tests that some plasma extends well above the disk midplane and scatters radiation along our line of sight \citep[for a review, see, {e.g.},][]{Parmar88}.
This region is referred to as the extended accretion disk corona \citep[hereafter ADC or extended disk atmosphere, see, {e.g.},][]{Psaradaki2018}; it is expected to be optically thin in X-rays and populated by highly ionized matter. X-ray spectroscopic studies of selected high-inclination WMNSs have confirmed the existence of such an extended ADC in \mbox{EXO 0748$-$676} \citep{Jimenez2003}, \mbox{4U 1822$-$37} \citep{Cottam2001}, or \mbox{2S 0921$-$63} \citep{Kallman2003}. 
The nature and geometry of ADCs are not well understood, but different models predict their existence \citep{White1982, Miller2000}.

X-ray polarimetry has introduced a novel diagnostic for NS LMXBs through two new observables, the polarization angle (PA) and the polarization degree (PD), and their variability. This new information can provide insight into the accretion physics of these objects and the geometry of the BL (and/or SL) and possibly of the ADC and the disk wind. \ixpe has already observed several systems. For \mbox{Cyg X-2}, the polarization angle (PA) was aligned along the position angle of the radio jet \citep{Farinelli23}, which is likely perpendicular to the accretion disk. On the other hand, \cite{LaMonaca2024} observed a significant difference between the PA measured from \mbox{Sco X-1} and the jet position angle, suggesting a more complex geometry for this source. In \mbox{Cir X-1} \citep{Rankin2024}, \mbox{XTE J1701$-$462} \citep{Cocchi2023} and \mbox{GX 5$-$1} \citep{Fabiani24} the polarization varied with the state and/or the hardness of the source with a PA rotation up to 67\degr\ in \mbox{Cir X-1} and PD passing from $\sim4\%$ in the horizontal branch to $\sim1\%$ in the Normal/Flaring branch of Z sources \citep{DiMarco24}. In the Atoll sources \mbox{GX 9+9} \citep{Ursini23}, \mbox{4U 1820$-$303} \citep{DiMarco23}, and \mbox{4U 1624$-$49} \citep{Saade24}, and the Z source \mbox{GX 340+0} \citep{Lamonaca24b} a dependence of the PD on energy was observed. 

\source is a persistent LMXB located at a distance of 7$\pm$1\,kpc \citep{Bandyo99}.
\cite{Fleischman85} observed for the first time Type-I X-ray bursts in this system, allowing unambiguous identification of its compact object as a NS.
It has been classified as a bright Atoll source \citep{Hasinger1989} or a Z source based on the strong secular evolution of its CCD and hardness intensity diagrams (HID) that showed peculiar features \citep{Giridharan2023}. For example in \cite{Stella1985}, the source HID shows two crossing branches, while \citet{Schnerr2003} reported a shape similar to Atoll sources but with a pattern opposite to the usual one observed in other WMNSs. Moreover, its fast variability properties, including an individual quasi-periodic oscillation at $\sim 61$ Hz, are not easily classified \citep{Homan98}. 
In general, identifying the source state(s) was found to be more difficult for \source than for other LMXBs.

\source belongs to the group of dipping sources \citep{dai2014, Iaria2014, Trigo10}. Its dips are reported to occur periodically every 24.5274(2) days \citep{Iaria2014} and are used to define the zero phase. The dip periodicity is compatible with the orbital period of the source reported by \cite{Corbet2010} on the basis of the modulation of the K-band and X-ray light curves. Off-phase X-ray dips were also occasionally observed \citep{Bobrikova24a,Trigo12}.

Spectral analyses of \source have been performed in the past based on data collected with different satellites, \cite[see, {e.g.},][and references therein]{Saavedra23}. In particular, the \xmm observations analyzed by \cite{Trigo12} showed a pronounced obscuration along the line of sight.
This is consistent with the presence of a disk wind and/or a warm disk atmosphere, provided the inclination of the system is in the 60\degr--80\degr\ range. Further analysis of the same \xmm data by \cite{Maiolino19} indicated an inclination ${\sim}$60\degr, whereas a ${\sim}$70\degr\ value was obtained from the \nustar data by \cite{Saavedra23}.

\ixpe conducted two observations of \source before the one presented in this article. The first on 2023 October 17--19 \citep{Bobrikova24a} and the second on 2024 February 25--27 \citep{Bobrikova24b}. In the first \ixpe observation, about one week before the periodic dip (orbital phase 0.74--0.83), \source exhibited peculiar characteristics. The source showed a continuous rotation of the PA by ${\sim}$70\degr\, together with a change in the dependence of PD on energy before and after a dip \citep{Bobrikova24a}. Correspondingly, the spectral properties showed little to no change. The second observation of \source was carried out simultaneously by \ixpe and \swift shortly after the end of the periodic dip (orbital phase 0.08--0.16). Polarization was measured with an almost constant PD of ${\sim}$2.5\% and PA of ${\sim}$24\degr\ \citep{Bobrikova24b}. The \swift spectrum of the source remained nearly constant during the \ixpe observation. 
\cite{Bobrikova24b} studied the differences and similarities between these two observations; the primary finding was that the second observation displayed a notably higher overall polarization, despite the almost identical state of the source. 

The present paper is based on \ixpe and coordinated observations of \source that were carried out in April 2024. Observations and data reduction are presented in  Sect.~\ref{sec:data}, the polarimetric and spectral analyses in Sect.~\ref{sec:analysis}. We discuss our results in Sect.~\ref{sec:discussion} and present conclusions in Sect.~\ref{sec:conclusion}.

\section{Observations and data reduction}\label{sec:data}

\begin{figure*}
    \centering
    \includegraphics[width=0.65\linewidth]{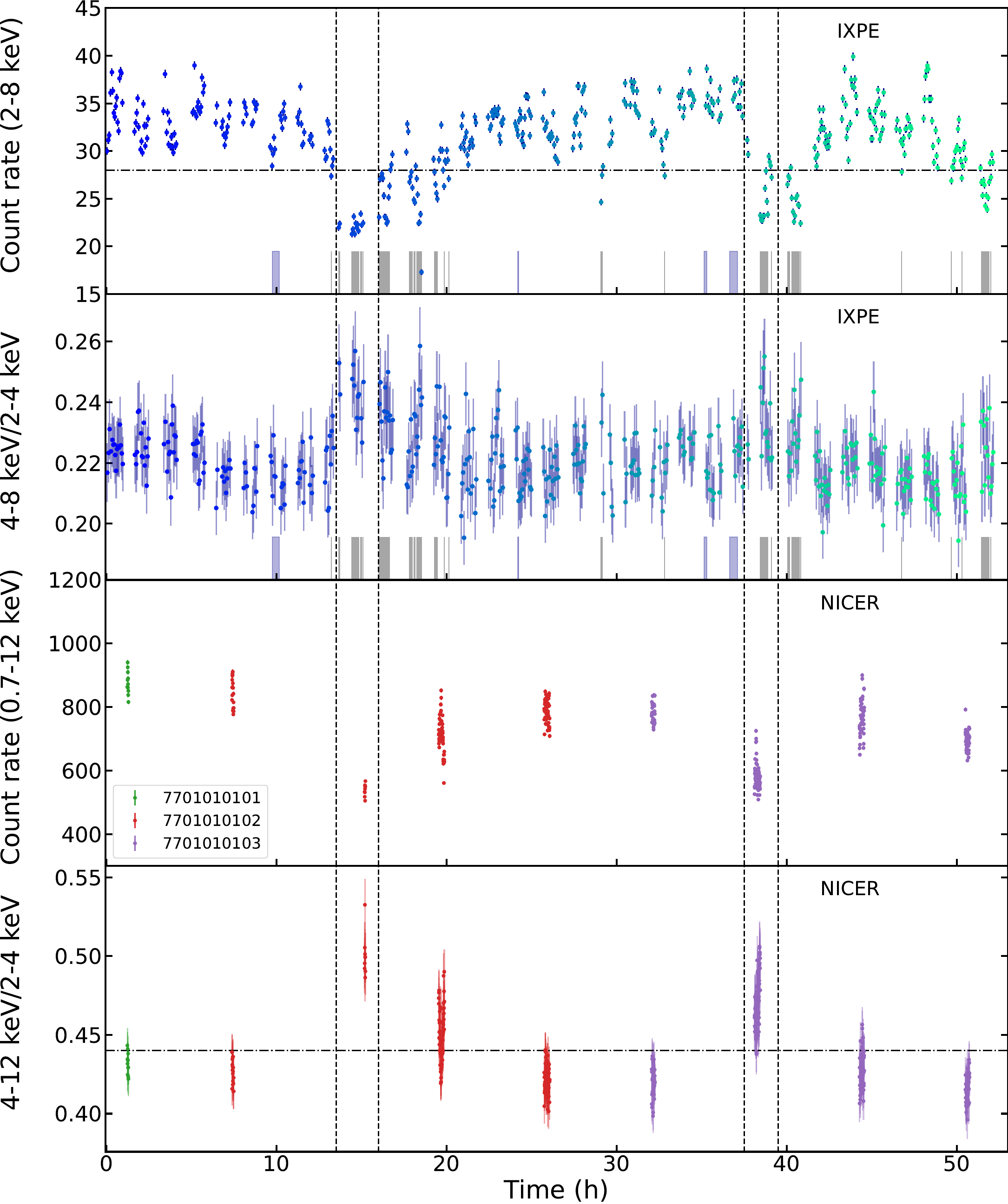}
    \caption{From top to bottom: Light curve of the third \ixpe observation of \source obtained in the 2--8~keV combining the three telescopes binned in 200~s and the corresponding hardness ratio as a function of time with the same binning; \nicer 20 s binned light curve in the 0.7--12~keV band, normalized to 50 detectors and the corresponding hardness ratio as a function of time. In the first two panels the light blue shaded bars mark the \swift observation periods, and the black horizontal dash-dotted line report the threshold applied to select the Dip state periods (corresponding to the gray vertical shaded bars) in the spectropolarimetric analysis (count rate ${<}28$ counts\,s$^{-1}$) with respect to Off-dip (count rate ${\geq}28$ counts\,s$^{-1}$). The vertical black dashed lines reported in these panels show the periods we applied to select the dips in the \texttt{pcube} model-independent analysis. In the fourth panel, the horizontal dash-dotted black line marks the threshold separating the Dip time intervals (hardness ratio ${>} 0.44$) from the Off-dip intervals (hardness ratio ${\leq} 0.44$) in the \nicer data.}
    \label{fig:lc_ixpe}
\end{figure*}

\begin{deluxetable*}{llcccc}
\tablecaption{Coordinated X-ray campaign during the third \ixpe observation of \source.}
\label{tab:exposure}
\tablehead{& Obs ID & Start (UTC) & Stop (UTC) & Telescope & Exp. time (ks)}
\startdata
\ixpe & 03003401 & 2024 April 20 20:49 & 2024 April 23 01:01 & DU 1 & 99.9 \\
& & & & DU 2 & 100.0\\
& & & & DU 3 &  100.1\\
\hline
\swift & 00036688052 & 2024 April 21 06:35 & 2024 April 21 06:58 & XRT-WT & 1.4 \\
& 00036688053 & 2024 April 21 21:01 & 2024 April 22 09:55 & XRT-WT & 2.4 \\
\hline
\nicer &  7701010101 & 2024 April 20 04:54 & 2024 April 20 22:06 & & 2.5 \\
& 7701010102 & 2024 April 21 04:08 & 2024 April 21 22:52 & & 3.0 \\
& 7701010103 & 2024 April 22 04:41 & 2024 April 22 23:33 & & 3.9 \\
\enddata
\end{deluxetable*}

\begin{figure}%[!t]
    \centering
    \includegraphics[width=\linewidth]{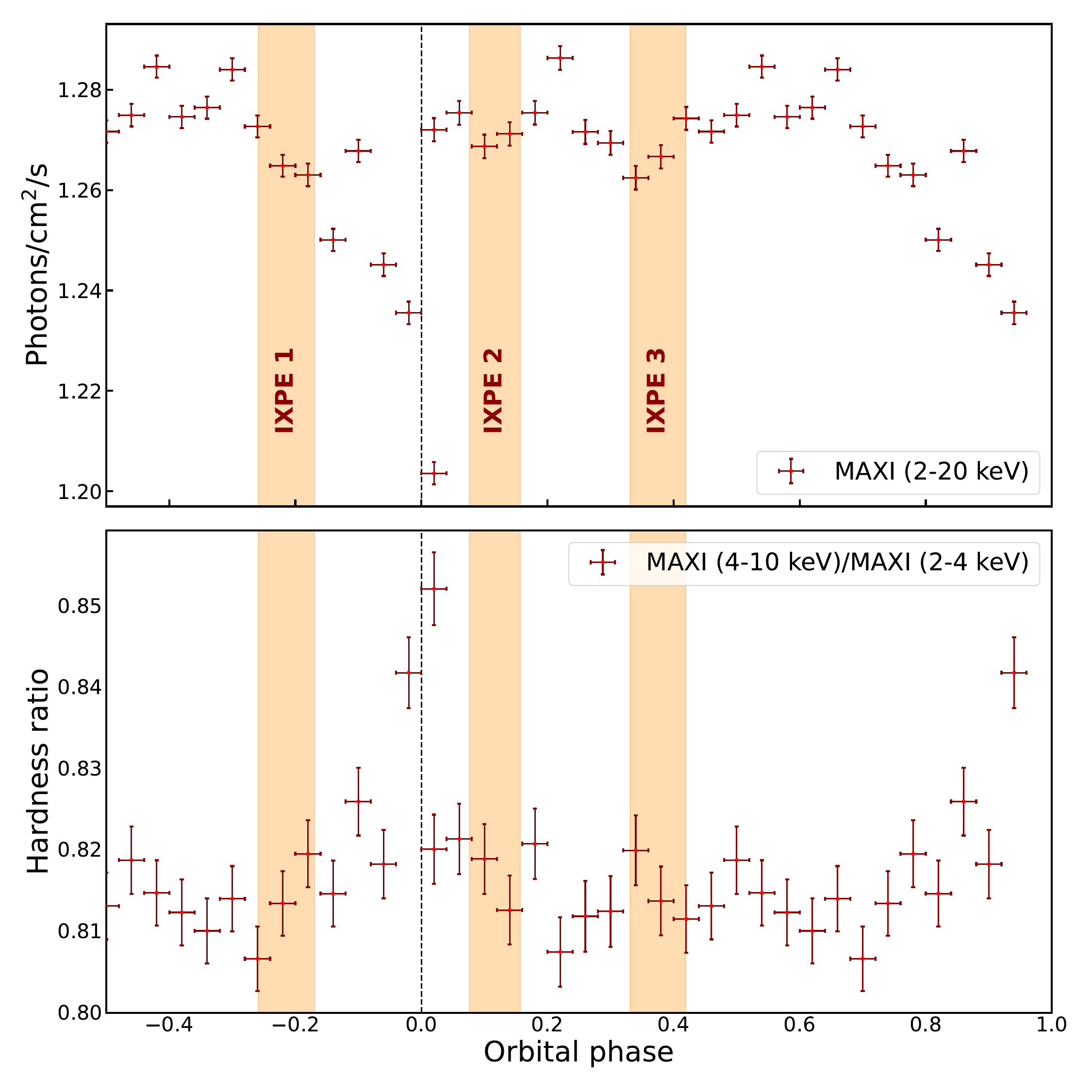}
    \caption{\maxi folded light curve and hardness ratio of \source using the ephemeris of \cite{Iaria2014}. The periodic dip defines zero phase. The orbital phase intervals covered in the three \ixpe observations (IXPE 1, 2, 3) are marked with vertical orange strips. It is apparent that the spectral hardness increases when the source count rate decreases.}
    \label{fig:maxi_folded}
\end{figure}

The \ixpe payload consists of three identical telescopes, providing imaging and spectral polarimetry over the 2--8~keV energy band \citep{Weisskopf2022,Soffitta_2021}. 
It observed \source for the third time (Observation ID 03003401) in the period 2024 April 20 20:49 UTC -- April 23 01:01 UTC (orbital phase 0.33--0.42), for a total effective exposure time of ${\sim}$100~ks for each telescope (see Table~\ref{tab:exposure}). The 2--8 keV \ixpe light curve binned every 200~s is shown in Figure~\ref{fig:lc_ixpe}, along with the hardness ratio of the rates in the 2--4 and 4--8 keV bands.

Two coordinated pointings, covering the 0.2--10 keV energy range, were carried out with the X-ray telescope (XRT) on board the Neil Gehrels Swift Observatory (\swift) \citep{Gehrels2004} to gather additional spectral information (see Table~\ref{tab:exposure}). The second pointing consisted of three different snapshots. All \swift observations were performed in Windowed Timing mode (WT).
The Neutron Star Interior Composition Explorer (\nicer) onboard the International Space Station (ISS) also pointed \source with its concentrator X-ray optics and silicon drift detectors operating in the 0.7--12 keV energy range \citep{Gendreau16}. \source observations were collected in three Observation IDs (see Table~\ref{tab:exposure}).\footnote{Observation ID 7701010101 covered only a very short interval of the \ixpe observation, so in the following spectral analysis we only considered the other two Observation IDs.}

\ixpe data were processed for polarimetric analysis using the \textsc{ixpeobssim} package version 31.0.1 \citep{Baldini2022}. For spectral and spectropolarimetric analysis, data extraction was performed using \textsc{HEASoft} version 6.33.2 and the standard \textsc{ftools} \citep{heasoft}. In the case of \ixpe, we used the CALDB released on 2024  February 28. Source events were extracted from \ixpe data by selecting a circular region of radius 100\arcsec\ centered on the position of the source, using \textsc{SAOImageDS9}.\footnote{\href{https://sites.google.com/cfa.harvard.edu/saoimageds9}{https://sites.google.com/cfa.harvard.edu/saoimageds9}} Due to the high flux of the source, the background was not subtracted, as prescribed by \cite{DiMarco_2023}.
In the model-independent polarimetric analysis, based on the \texttt{pcube} algorithm in \textsc{ixpeobssim}, the unweighted analysis was adopted, whereas for the model-dependent spectral and spectropolarimetric analyses we used the `weighted analysis'. The latter is based on the ellipticity of the photoelectrons tracks collected by the \ixpe detector units (DUs) and follows the approach of \citet{DiMarco_2022}. Data were binned so as to have at least 30 counts per energy channel in the spectral analysis, while in the spectropolarimetric analysis we applied a constant energy binning of 120~eV for all three energy distributions of the Stokes parameters $I$, $Q$, and $U$, given the lower statistic available in the $Q$ and $U$ spectra.

The extraction of source and background events was performed in the \swift data with \textsc{SAOImageDS9} using an annulus (inner radius of about 5\arcsec\ and an outer radius of about 40\arcsec) in order to avoid possible pile-up effects \citep{Romano2006}; the source annulus was centered at the \source coordinates, and the background annulus about 80\arcsec\ off-source. The \swift data were fitted in the 0.8--8~keV band to optimize the statistics and grouped so as to have at least 30 counts per bin. The response matrices released in the HEASARC CALDB on 2023 July 25 were applied in the analysis.

\nicer data were processed with the \nicer Data Analysis Software v012a released on 2024 February 9 and with the CALDB version released on 2024 February 27. The background spectra were estimated by applying the \textsc{SCORPEON} model. All spectra, fitted in the 1.5--10~keV energy band, were grouped so as to have at least 30 counts per bin.
Light curves, HID, and time selections were obtained using \textsc{stingray} \citep{stingray1,stingray2,Bachetti24}.

\section{Data analysis}\label{sec:analysis}

The \ixpe light curve of the third observation of \source displayed two erratic, non-periodic dips, during which the source count rate decreased by 20--30\% and the hardness ratio increased (see Figure~\ref{fig:lc_ixpe}). The first lasted about 12~ks (including ingress, dip phase, and egress), the second ${\sim}5$~ks. Both dips showed fast continuous ingress, while the egress was characterized by marked variability. Their orbital phase was 0.334 and 0.395, thus away from the periodic dip that defines the zero phase (see Figure~\ref{fig:maxi_folded} which also displays the orbital phase intervals covered by the other \ixpe observations).

\begin{figure}%[!h]
    \centering
    \includegraphics[width=\linewidth]{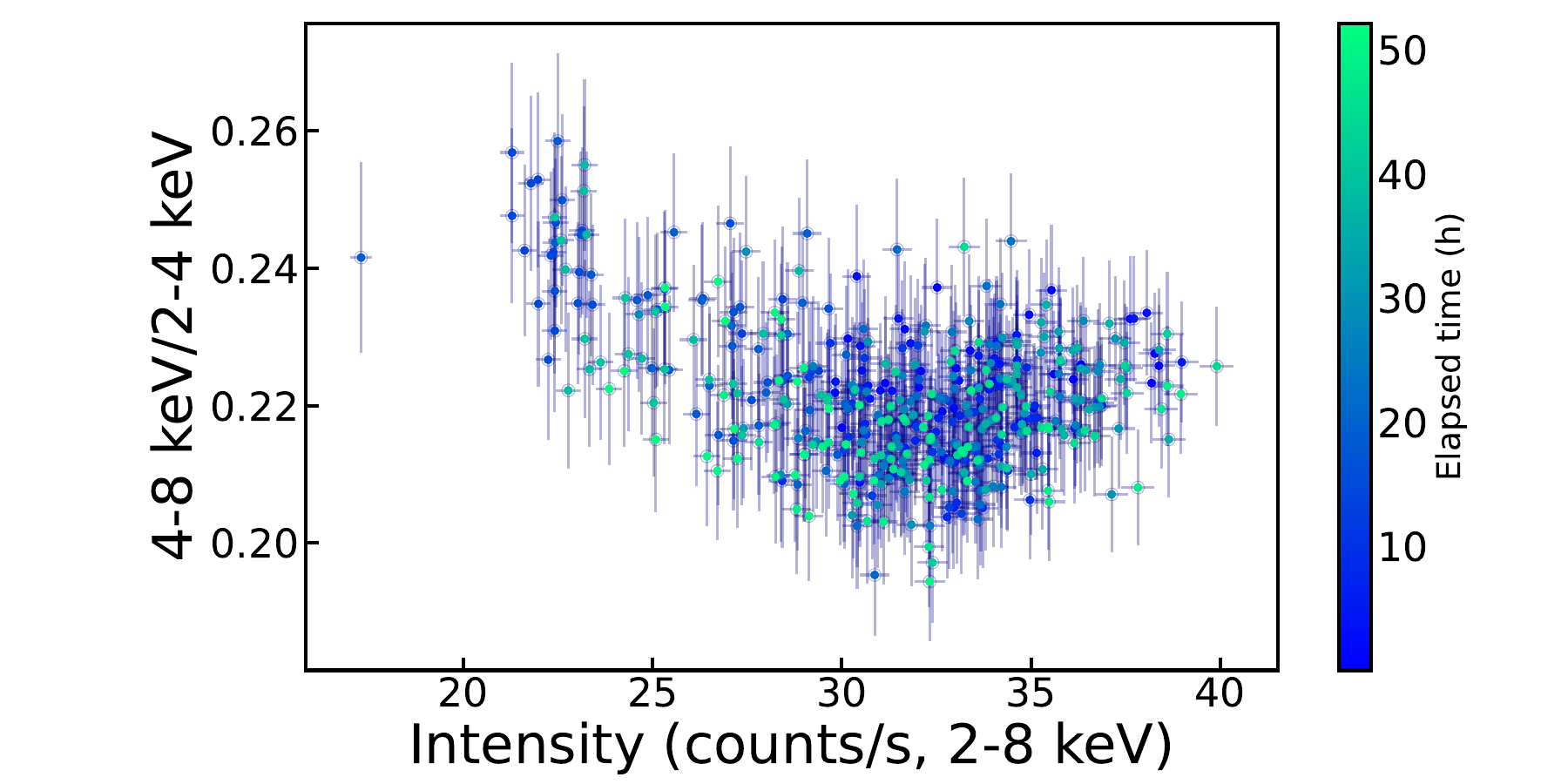}
    \caption{Hardness-intensity diagram obtained from \ixpe data, using the same energy bands (4--8\,keV and 2--4\,keV) and the same binning (200~s) as in \cite{Bobrikova24a,Bobrikova24b}. The colors follow the same time behavior as in  Figure~\ref{fig:lc_ixpe} and the colorbar is on the right.}
    \label{fig:hid}
\end{figure}

The hardness intensity diagram obtained from the \ixpe data is shown in Figure~\ref{fig:hid}. \nicer data partially covered the two dips, as shown in Figure~\ref{fig:lc_ixpe}, thus making detailed spectral modeling possible. No \swift data were taken during the dips.

\subsection{Polarimetric analysis}

\begin{figure}%[!h]
\includegraphics[width=\linewidth]{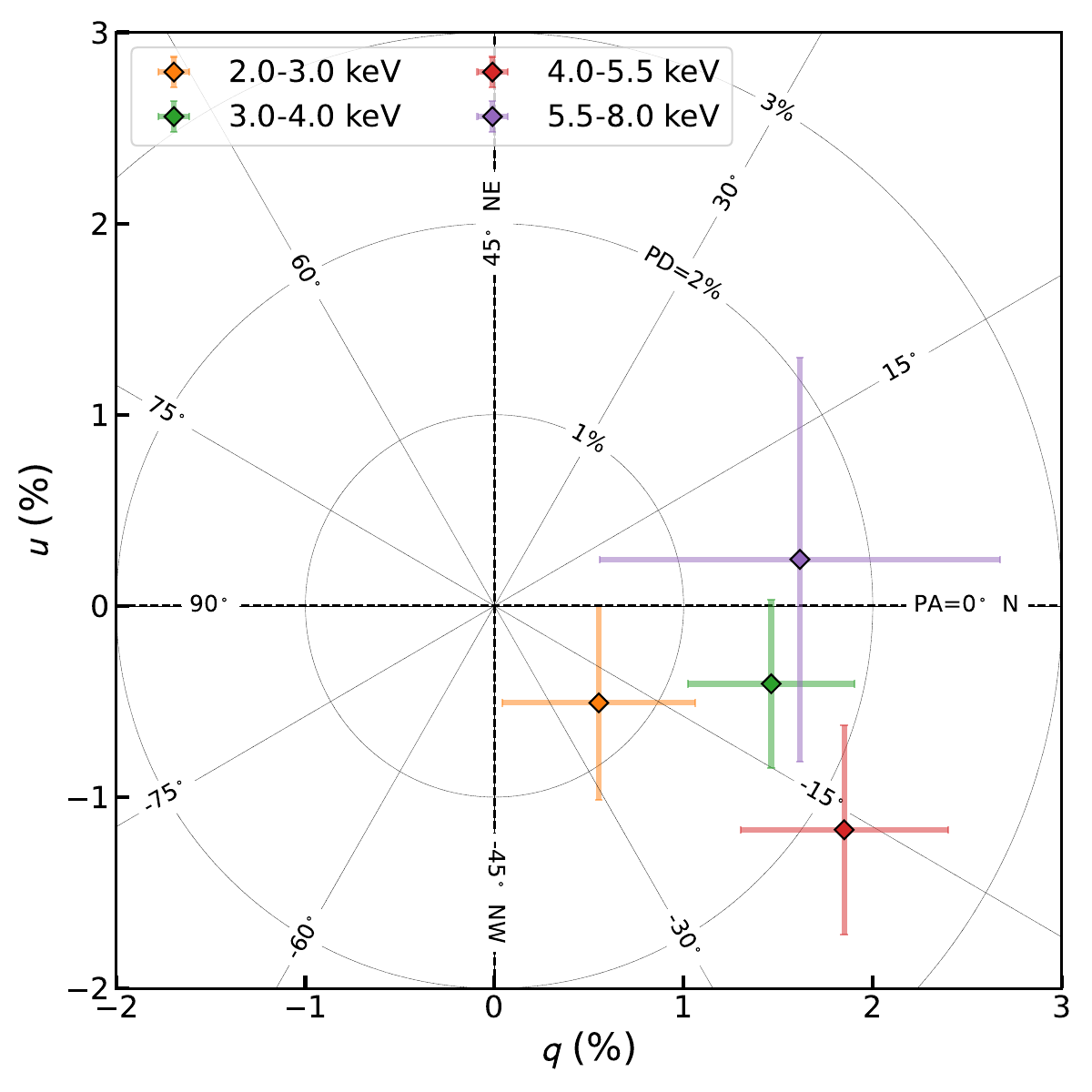}
\caption{Normalized Stokes parameters $q=Q/I$ and $u=U/I$ in different energy bins as obtained by \texttt{pcube} in \textsc{ixpeobssim}. The error bars are at 68\% CL. No significant trend of the polarization with energy is observed.}
\label{fig:q_u_plane_energy}
\end{figure}

\ixpe data in the 2--8\,keV band were used in the polarimetric analysis. We obtained an average polarization of PD=$1.4\%\pm0.3\%$ with the PA of $-11\degr \pm 7 \degr$ (hereafter errors are at 68\% confidence level, unless stated otherwise). A search for energy trends in the polarization was carried out. The results, shown in Figure~\ref{fig:q_u_plane_energy}, are compatible with an energy-independent polarization, to within 2$\sigma$.

\begin{figure*}%[!t]
\centering
\includegraphics[width=0.45\linewidth]{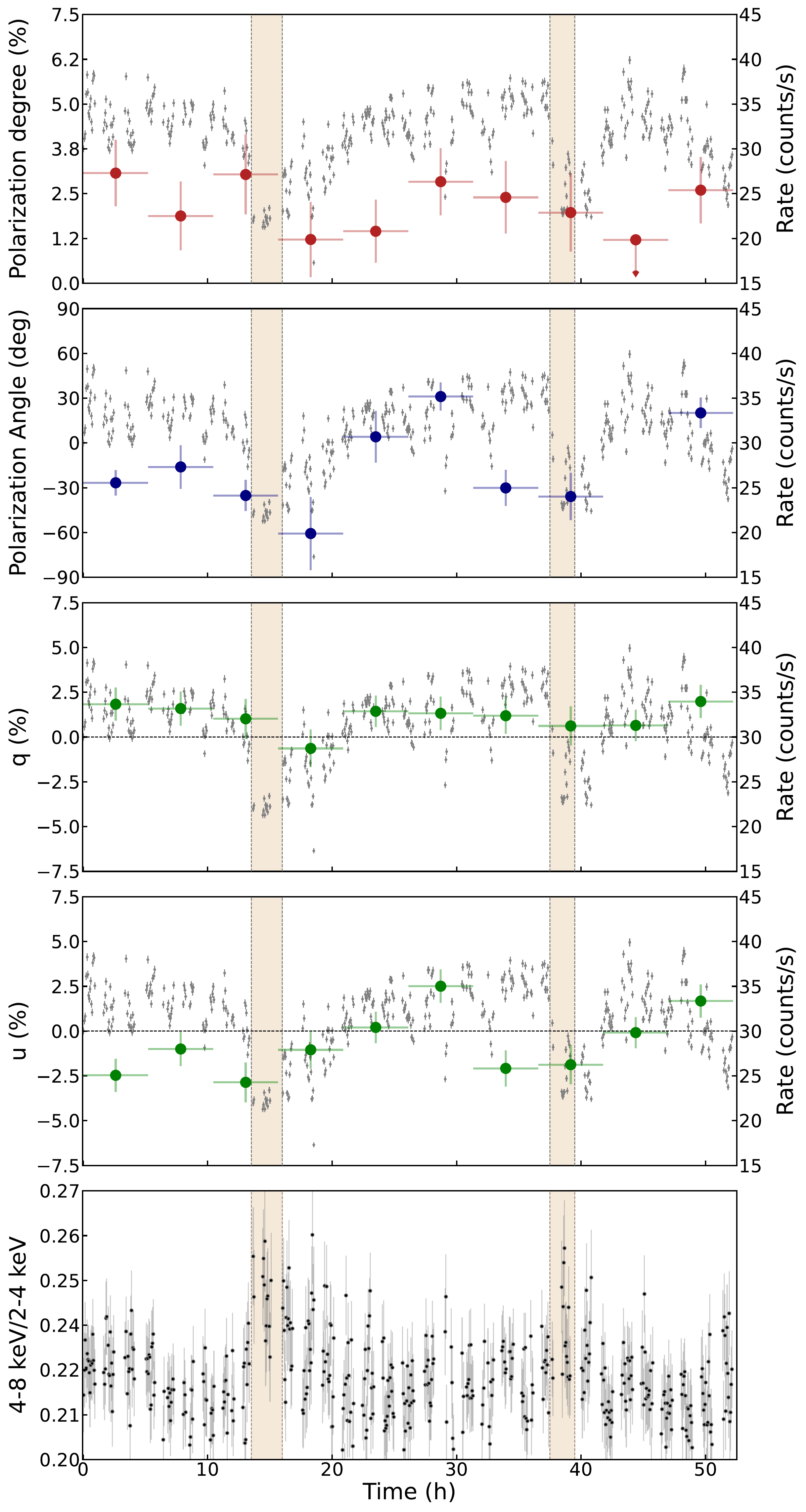}
\includegraphics[width=0.45\linewidth]{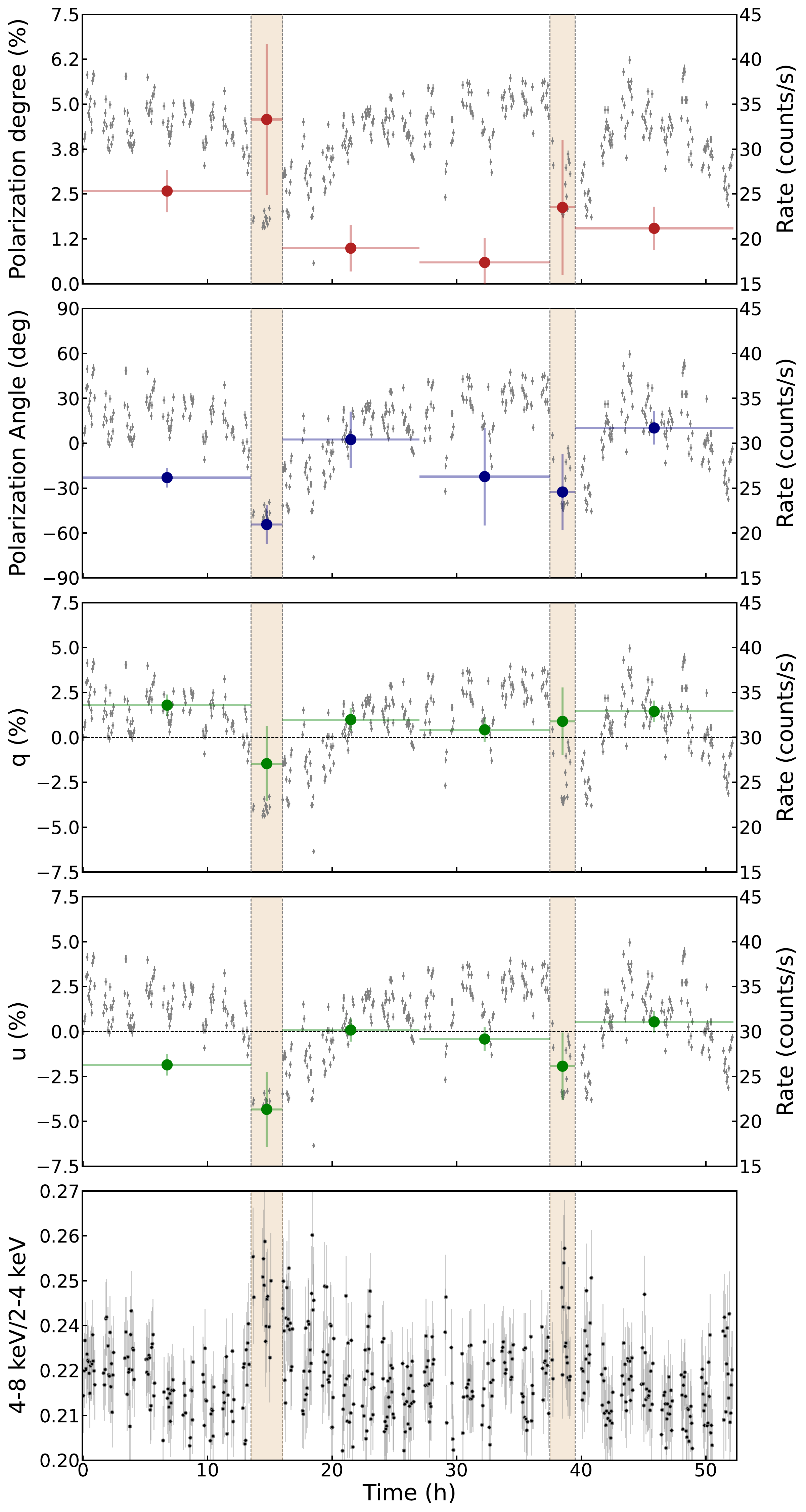}
\caption{Polarization properties of \source as a function of time during the third \ixpe observation. The panels from top to bottom show PD, PA, $q=Q/I$, $u=U/I$ and the hardness ratio along the observation. The gray points in the top four panels show the 2--8 keV count rate (right axis). Left panels: a constant binning of ${\sim}5.2$~hours is adopted. Right panels: the observation is divided into six time bins, two of them corresponding to the dips, and the other four covering intervals before, after, and between the two dips. Errors are at 68\% CL.}
\label{fig:time_pol}
\end{figure*}

We investigated polarization properties on time scales shorter than the observation duration, by using ${\sim}$5.2~hr-long intervals. The results in Figure~\ref{fig:time_pol}-left show a marked variability in the PA, which evolves in time in a non-continuous way. We note that during the first \ixpe observation of \source the PA displayed instead a continuous rotation in time \citep{Bobrikova24a}.

A further time resolved analysis was performed by dividing the observation into six intervals, two during the dips and four outside of the dips. The polarimetric behavior with this time selection is reported in Figure~\ref{fig:time_pol}-right. It shows a rotation of PA from north to west, likely associated with an increase of PD during the dips. A similar behavior was observed during the dip in the first \ixpe observation \citep{Bobrikova24a}. Note that the second dip, which is shorter and less pronounced, has larger polarimetric uncertainties. Different time selections were attempted, but narrow selections result in a not significant PD, while larger time selections report almost the same results, except when the selection starts to include periods with higher count rate.

We also performed the energy-resolved polarimetric analysis in each time interval of Figure~\ref{fig:time_pol}-right, but owing to the lower statistics no clear conclusion could be drawn. We note that Figure~5 of \cite{Bobrikova24a} provided marginal evidence for polarimetric variability with energy across the dip of the first observation. 

\subsection{Spectral analysis}

\begin{deluxetable*}{ccccccccc}%[!b]
\tablecaption{Best-fit parameters for the spectral model \texttt{tbabs*(diskbb+bbodyrad)} as obtained from the two \swift spectra in the 0.8--8.0~keV energy band. Errors are reported at 68\% CL.}
\label{tab:spectrum}
\tablehead{ObsID & \texttt{tbabs} & \multicolumn{2}{c}{\texttt{diskbb}} & \multicolumn{2}{c}{\texttt{bbodyrad}} & $\chi^{2}$/d.o.f. & Flux$_{2-8\,\text{keV}}$  & Hardness \\
& $N_{\text{H}}$ (10$^{22}$ cm$^{-2}$) & $kT_{\text{in}}$\tablenotemark{a} (keV) & norm & $kT_{\text{bb}}$ (keV) & norm & & (10$^{-9}$ erg~s$^{-1}$~cm$^{-2}$) &}
\startdata
00036688052 & $4.63\pm0.10$ & [1.0] & $370\pm20$ & $1.38^{+0.04}_{-0.03}$ & $150\pm20$ & 610/568 & 6.0 & 0.30 \\
00036688053 & $4.60\pm0.07$ & [1.0] & $360\pm20$ & $1.46\pm0.02$ & $160\pm12$ & 660/638 & 6.8 & 0.35\\ %\hline
\enddata
\tablenotetext{a}{Fixed.}
\end{deluxetable*}

During the third \ixpe observation coordinated \swift and \nicer observations were also available (see Sect.~\ref{sec:data}) and provided additional spectral coverage in the soft X-ray band. Therefore, we adopted the same spectral model introduced by \cite{Trigo12} to model \xmm data from \source. In the spectral and spectropolarimetric analyses, we used \textsc{xspec} v.12.14.0 \citep{Arnaud96}. The continuum was modeled by the sum of a soft multicolor disk, \textit{i.e.} \texttt{diskbb} \citep{Mitsuda84}, describing the emission from the inner disk and/or the NS surface, and a harder Comptonized emission in an optically thick environment such as SL/BL, approximated here with \texttt{bbodyrad}. Absorption was modeled with \texttt{TBabs}, which is based on the abundances reported in \cite{Wilms2000}. Our overall continuum model was thus \texttt{tbabs*(diskbb+bbodyrad)}.
The two \swift observations, which were both away from the dips, showed consistent spectra, with best-fit parameters also in agreement with those reported in \cite{Bobrikova24b} (see Table~\ref{tab:spectrum}).

\begin{figure}%[!h]
\centering
\includegraphics[width=0.85\linewidth]{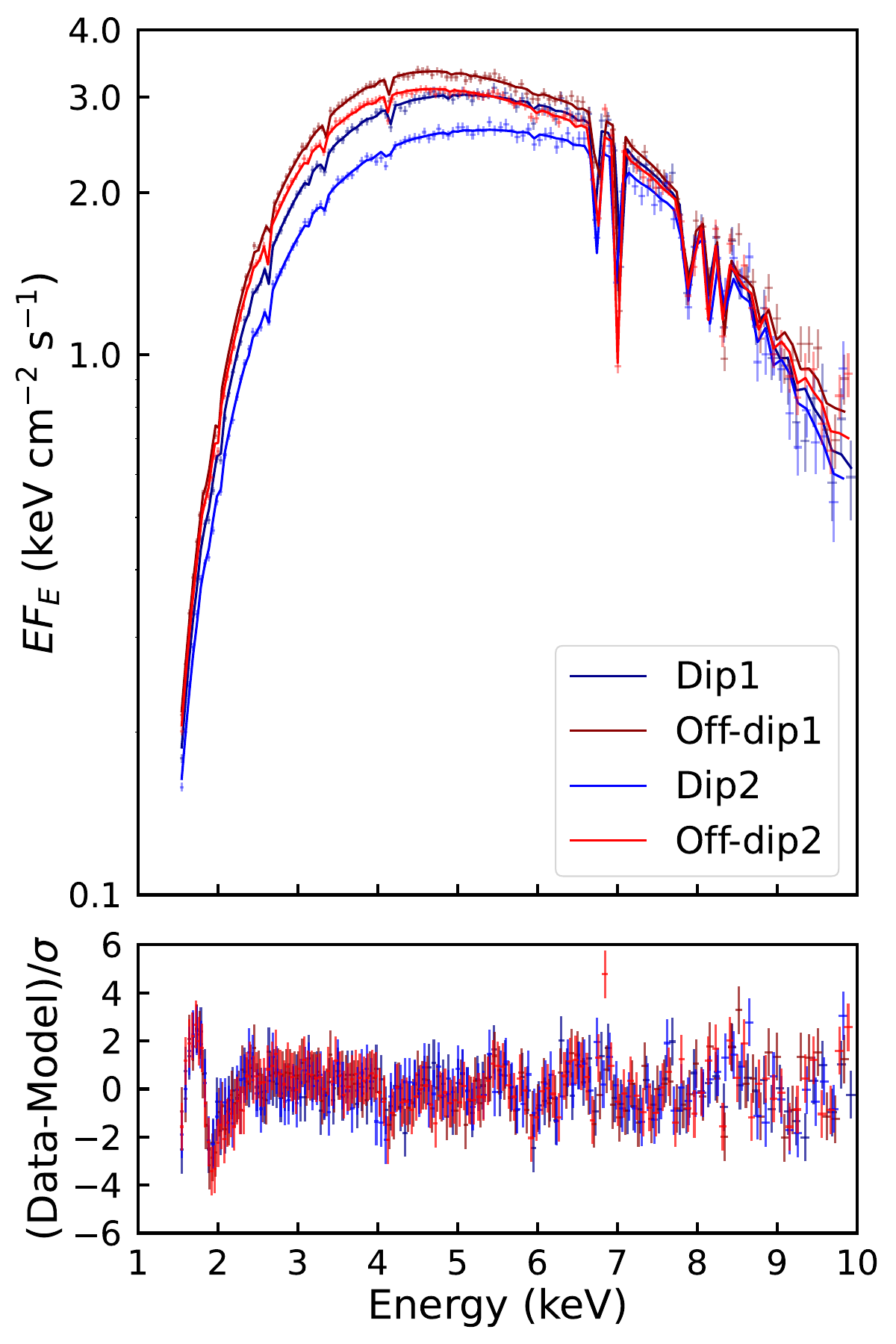}
\caption{Best-fit spectral fits from \nicer during Dip and Off-dip states in Observation IDs 77010102 and 77010103.}
\label{fig:nicer_spec}
\end{figure}

The \nicer data partially covered the \ixpe observation with several snapshots, some of which were during dips (see Table~\ref{tab:exposure} and Figure~\ref{fig:lc_ixpe}). To model the spectra in the Dip and Off-dip states, the data were selected on the basis of their hardness ratio (HR), as shown in Figure~\ref{fig:lc_ixpe}. By virtue of its better spectral capabilities and throughput (compared to \ixpe and \swift) \nicer resolved some of the absorption features reported in the literature \citep{Trigo12,dai2014,Saavedra23}. In particular, residuals in the 6-8 keV region were clearly visible when the continuum model was fitted to the spectra. To account for those features, we followed the approach in \cite{Trigo12} and 
included a \texttt{warmabs} component to model absorption by photoionized plasma along the line of sight. 
Moreover, a \texttt{cabs} component was included to account for the effects of Thomson/Compton scattering, which was found to be significant when modeling both the persistent and dipping emission \citep{Boirin05,Trigo06,Trigo12}. The latter component was forced to have $\times 1.21$ the column density of the \texttt{warmabs} component \citep[which accounts for the number of electrons per hydrogen atom for solar abundances material][]{Stelzer99}. Furthermore, a \texttt{gaussian} component was included to account for a broad Fe line emission. The complete model we fit to the \nicer spectra was thus \texttt{TBabs*warmabs*cabs*(diskbb+bbodyrad+gauss)}. The best-fit results reported in Table~\ref{tab:spectrum2} are in general agreement with those from \xmm and \chandra observations \citep{Trigo12,dai2014}. The spectra are shown in Figure~\ref{fig:nicer_spec}. Absorption was higher during the dips, especially during the first one, whereas the  \texttt{diskbb} temperature was lower, ${\sim}0.5$~keV, than outside the dips (${\sim}1$~keV).
The \texttt{bbodyrad} component remained instead at a nearly constant temperature of ${\sim}1.1-1.2 $~keV.

\begin{deluxetable*}{crcccc}[!thb]
\tablecaption{Best-fit parameters for the spectral model \texttt{tbabs*cabs*warmabs*(diskbb+bbodyrad+gauss)} as obtained from different \nicer spectra in the 1.5--10.0~keV energy band. Errors are reported at 68\% CL.}
\label{tab:spectrum2}
\tablehead{ Observation ID & & \multicolumn{2}{c}{7701010102} & \multicolumn{2}{c}{7701010103} \\
State &  & Dip & Off-dip & Dip & Off-dip}
\startdata
\texttt{tbabs} & $N_{\text{H}}$ (10$^{22}$ cm$^{-2}$) & $3.75_{-0.02}^{+0.10}$ & $3.61_{-0.01}^{+0.07}$ & $3.69_{-0.02}^{+0.01}$ & $3.58^{+0.07}_{-0.01}$ \\
\texttt{cabs}\tablenotemark{a} & $N_{\text{H}}$ (10$^{22}$ cm$^{-2}$) & [61] & [48] & [40] & [47] \\
\texttt{warmabs}\tablenotemark{b} & $N_{\text{H}}$ (10$^{22}$ cm$^{-2}$) & $51_{-2}^{+3}$ & $40_{-1}^{+2}$ & $33\pm2$ & $39\pm2$ \\
& $r\log \xi$ & $3.58_{-0.06}^{+0.01}$ & $3.52_{-0.02}^{+0.05}$ & $3.53_{-0.02}^{+0.08}$ & $3.50_{-0.02}^{+0.01}$ \\
%& Si abund & [1.3]\tablenotemark{b} & $1.9_{-0.8}^{+1.4}$ & [1.3]\tablenotemark{b} & $3.7\pm1.6$ \\
& $v_{\rm turb}$ (km/s) & $117_{-13}^{+40}$ & $271_{-4}^{+26}$ & $210\pm40$ & $274_{-7}^{+30}$ \\
& velocity (km/s) & $-2410\pm30$ & $-2332\pm15$ & $-1951_{-20}^{+7}$ & $-1962\pm15$ \\
\texttt{diskbb} & $kT_{\rm in}$ (keV) & $0.52\pm0.14$ & $1.02\pm0.06$ & $0.52\pm0.03$ & $1.09\pm0.01$ \\
& norm & $2450_{-30}^{+50}$ & $237\pm2$ & $1726_{-20}^{+30}$ & $174_{-2}^{+690}$ \\
& $R_{\rm in}$\tablenotemark{c} (km) & $59.2_{-0.7}^{+1.2}$ & $18.4_{-0.2}^{+0.9}$ & $49.7_{-0.6}^{+0.9}$ & $15.8^{+63.8}_{-0.2}$ \\
\texttt{bbodyrad} & $kT_{\rm bb}$ (keV) & $1.08\pm0.07$ & $1.20\pm0.06$ & $1.11\pm0.02$ & $1.17\pm0.02$ \\
& norm & $773_{-50}^{+3}$ & $432\pm2$ & $498_{-2}^{+43}$ & $420_{-2}^{+40}$ \\
& $R_{\rm bb}$\tablenotemark{c} (km) & $19.4^{+0.1}_{-1.3}$ & $14.5\pm0.1$ & $15.6_{-0.1}^{+1.3}$ & $14.3_{-0.1}^{+1.4}$\\
\texttt{gauss} & $E$ (keV) & $6.55\pm0.16$ & $6.92_{-0.08}^{+0.06}$ & $6.58^{+0.05}_{-0.17}$ & $6.93_{-0.06}^{+0.04}$ \\
& $\sigma$ (keV) & $1.4\pm0.3$ & $0.86_{-0.05}^{+0.07}$ & $1.48\pm0.04$ & $1.24_{-0.04}^{+0.05}$ \\
& norm ($10^{-2}$)& $9.5_{-0.3}^{+0.2}$ & $1.96_{-0.08}^{+0.17}$ & $7.8^{+0.2}_{-0.5}$ & $4.0_{-0.9}^{+0.1}$ \\
\textbf{$\chi^2$/d.o.f.}\tablenotemark{d} & & 113/112 & 142/118 & 117/111 & 169/121 \\\hline
\multicolumn{2}{r}{Flux$_{2-8\text{ keV}}$ ($10^{-9}$ erg $\textrm{s}^{-1} \textrm{cm}^{-2}$)} & 4.95 & 5.50 & 4.26 & 5.14 \\
\enddata
\tablenotetext{a}{Fixed at 1.21 times the $N_{\rm H}$ value of \texttt{warmabs}.}
\tablenotetext{b}{The Ni abundance was frozen at 5 to fit of an absorption feature at ${\sim}$7~keV; this value is in line with that reported by \cite{Saavedra23}.}
\tablenotetext{c}{The inner radius for the \texttt{diskbb} component is estimated assuming an inclination at 70\degr\  \citep{Trigo12,dai2014,Saavedra23} and a distance of 7~kpc \citep{Bandyo99}. For $R_{\rm bb}$ in the \texttt{bbodyrad} component, the same value for the distance was used.}
\tablenotetext{d}{The higher $\chi^2$ values at higher flux is probably related to an instrumental effect at low energy close to the Si edge at ${\sim}$1.84~keV.}
\end{deluxetable*}

\subsection{Spectropolarimetric analysis}

In the spectropolarimetric analysis, we divided the \ixpe data from the third observation into Dip and Off-dip states, by using an intensity threshold of 28 cts\,s$^{-1}$ (see the horizontal dashed-dotted line in the top panel of Figure~\ref{fig:lc_ixpe}). Owing to the lower throughput and spectral capabilities of \ixpe (compared to those of \nicer) our spectropolarimetric analysis resorted to a simplified model, \texttt{tbabs*(diskbb+bbodyrad+gauss)}. The temperatures of both the \texttt{diskbb} and \texttt{bbodyrad} components, the energy and width of the Gaussian describing the iron line, were frozen at the values obtained from the \nicer spectral fits. 

As a first attempt, we assumed the same constant polarization for \texttt{diskbb+bbodyrad} and no polarization for the Gaussian \citep[see, {e.g.},][]{Churazov2002}: \texttt{tbabs*(polconst*(diskbb+bbodyrad)+gauss)}. The best-fit from the Dip data gave\footnote{The other free parameters in the fit --- such as the cross-normalizations, norm of \texttt{diskbb}, \texttt{bbodyrad} and \texttt{gaussian} --- in both Dip and Off-dip cases have values compatible with the ones reported in Table~\ref{tab:spec_pol}.} $\textrm{PD}=2.5\% \pm 0.7\%$ and $\textrm{PA}=-37\degr \pm 8\degr$ ($\chi^2/{\rm d.o.f.}=441/433$). For the Off-dip data the best-fit yielded $\textrm{PD}=1.2\% \pm 0.3\%$ with $\textrm{PA}=-7\degr\pm7\degr$ ($\chi^2/{\rm d.o.f.}=469/433$). These results parallel those obtained from the model-independent analysis: during the dips, when absorption was higher, the degree of polarization was higher, and the polarization angle was different with respect to the Off-dip state.

As a second step, we associated separate \texttt{polconst} polarization components to the \texttt{diskbb} and \texttt{bbodyrad} components: \texttt{tbabs*(polconst*diskbb+polconst*\\bbodyrad+gauss)}. 
The best-fit results for this model are reported in Table~\ref{tab:spec_pol}. Polarization was detected at a level of $\textrm{PD}{\sim} 4 \% $across the Off-dip and Dip states in the harder component (\texttt{bbodyrad}), together with marginal evidence for a PA change of $\sim 20\degr$. For the \texttt{diskbb} component we found $\textrm{PD}=10\%\pm4\%$ in the Dip state, whereas in the Off-dip state we derived an upper limit of $\textrm{PD}{<}4.3\%$ at 90\% CL. We warn that the soft component's 10\% PD during the dip should be considered with care, as the polarizations of the two components in the fit result are almost orthogonal, meaning that they subtract from each other and partial degeneracy sets in.

\begin{deluxetable}{crcccc}
\tablecaption{Best-fit parameters for the spectropolarimetric model \texttt{tbabs*(polconst*diskbb+polconst*bbodyrad+gauss)}. The unabsorbed flux is $6.4\times10^{-9}$ erg\,s$^{-1}$\,cm$^{-2}$
in the 2--8~keV energy range, corresponding to a luminosity of $3.8\times10^{37}$erg/s. Errors are at 68\% CL.}
\label{tab:spec_pol}
\tablehead{\multicolumn{2}{r}{State} & Dip & Off-dip}
\startdata
\texttt{tbabs} & $N_{\text{H}}$ (10$^{22}$ cm$^{-2}$) & $5.9\pm0.3$ & $4.40 \pm 0.05$ \\
\texttt{diskbb} & $kT_{\rm in}$ (keV) & [0.5] & [1.0] \\
& norm & $5600\pm600$ & $241 \pm 5$ \\
& $R_{\rm in}$\tablenotemark{a} (km) & $90\pm9$ & $18.6 \pm 0.4$ \\
\texttt{polconst1} & PD (\%) & $10\pm4$ & $<4.3$\tablenotemark{b} \\
& PA (deg) & $64\pm13$ &  -- \\
\texttt{bbodyrad} & $kT_{\rm bb}$ (keV) & [1.1] & [1.2] \\
& norm & $345\pm2$ & $223\pm2$ \\
& $R_{\rm bb}$\tablenotemark{a} (km) & $13.00\pm0.08$ & $10.5\pm0.1$ \\
\texttt{polconst2} & PD (\%) & $5\pm2$ & $3.1 \pm 1.1$ \\
& PA (deg) & $-32\pm7$ & $-8\pm10$ \\
\texttt{gauss} & $E$ (keV) & [6.6] & [6.9] \\
& $\sigma$ (keV) & [1.4] & [1.0] \\
& norm ($10^{-2}$)& $3.3\pm0.2$ & $2.1\pm0.2$ \\
& $f_{\rm DU1}$ & [1] & [1] \\
& $f_{\rm DU2}$ & $1.024 \pm 0.005$ & $1.021 \pm 0.002$ \\
& $f_{\rm DU3}$ & $1.000 \pm 0.005$ & $1.002 \pm 0.002$ \\
\multicolumn{2}{r}{$\chi^2$/d.o.f.} & 433/431 & 465/431 \\ \hline
\multicolumn{2}{r}{Flux$_{2-8\text{ keV}}$ ($10^{-9}$ erg\,s$^{-1}$\,cm$^{-2}$}) & 3.7 & 4.8 \\% \hline
\multicolumn{2}{r}{Flux$_{\texttt{diskbb}}$/Flux$_{\texttt{bbodyrad}}$} & 0.2 & 0.5 \\% \hline
\enddata
\tablenotetext{a}{The inner radius for the \texttt{diskbb} component was estimated assuming an inclination at 70\degr\  \citep{Trigo12,dai2014,Saavedra23} and a distance of 7~kpc \citep{Bandyo99}. For $R_{\rm bb}$ in \texttt{bbodyrad} component and the unabsorbed luminosity the same value of the distance was used.}
\tablenotetext{b}{Upper limit at 90\% CL.}
\end{deluxetable}

\section{Discussion}\label{sec:discussion}

Polarization during the third \ixpe observation of \source showed marked variability. The behavior of the PA over time, shown in Figure~\ref{fig:time_pol}, was complex and somewhat different from the continuous rotation seen during the first observation \citep{Bobrikova24a}. Comparing this behavior with the trend of the hardness ratio reported in the bottom panels suggests that PD increased when HR was higher and that PA increased at higher count rates. The right panels of Figure~\ref{fig:time_pol} also show that the polarization parameters in the two dips are compatible with each other, to within the uncertainties. 

\begin{figure*}%[!b]
\centering
\includegraphics[width=0.49\linewidth]{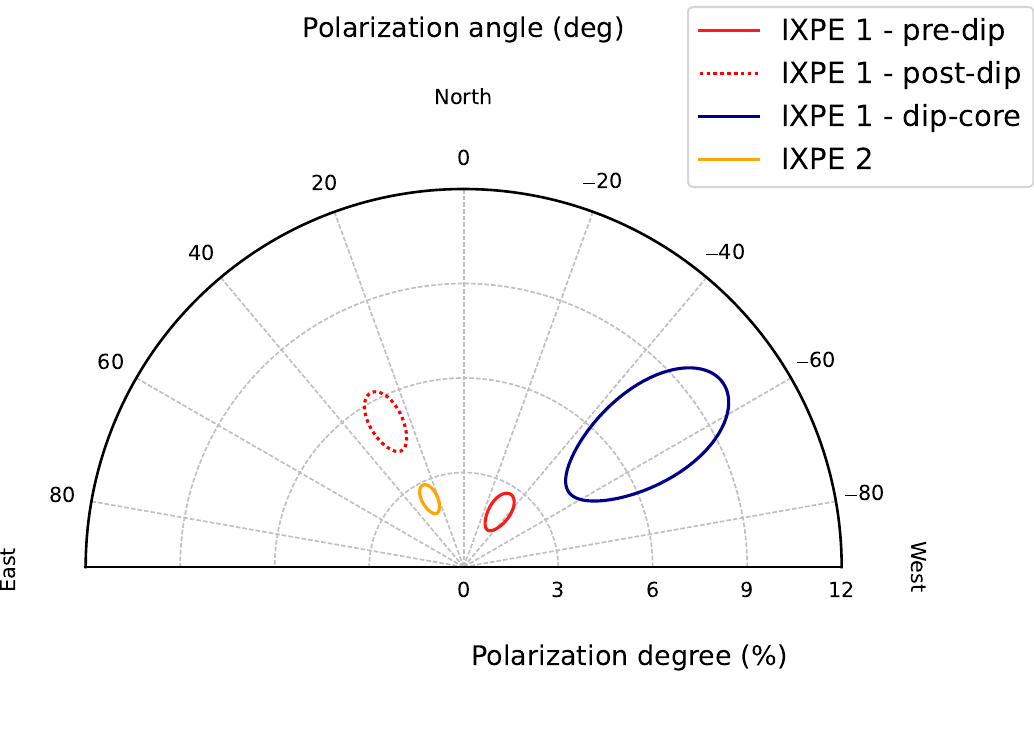}
\includegraphics[width=0.49\linewidth]{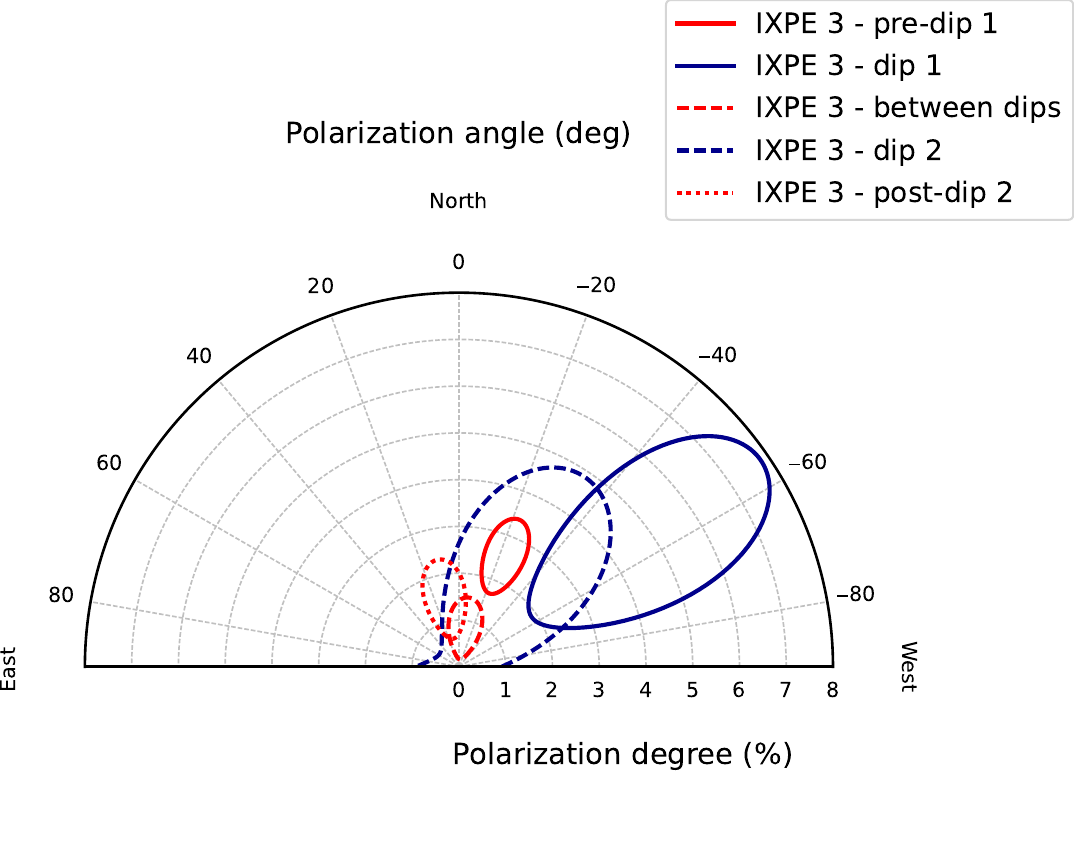}
\caption{Comparison between the polarization in the first two \ixpe observations of \source (left), and this new one (right) to compare the polarization of the dips and of the source in Off-dip states. Contours are at 90\% CL.}
\label{fig:comparison}
\end{figure*}

In Figure~\ref{fig:comparison}, we compare the evolution of the polarization properties during the third and the previous two observations. It is seen that the PA in the two pre-dip intervals of the third observation was similar to that of the first observation's pre-dip. The PA during the two dips was compatible with the dip PA in the first observation. Then, after the second dip, the PA attained a value similar to the post-dip PA of the first observation, which in turn is compatible with the PA during the entire second \ixpe observation (which did not show dips). We conclude that the polarimetric properties evolved in a similar manner across the three observations. 

\begin{figure}%[!h]
\centering
\includegraphics[width=0.8\linewidth]{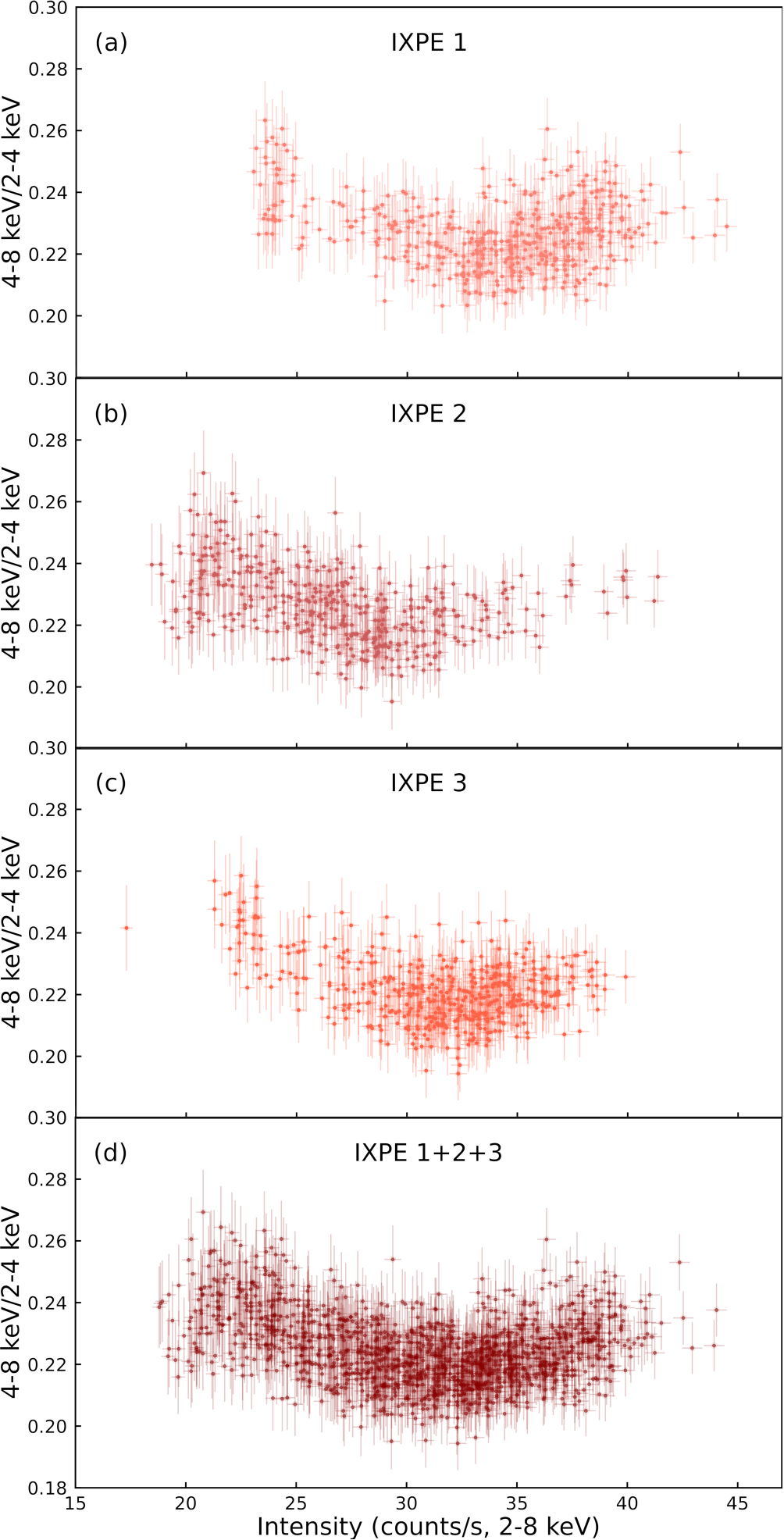}
\caption{\ixpe HID obtained for \source from the different observations (panels a--c) and when the three observations are combined (d).}
\label{fig:combined_hid}
\end{figure}

Concerning spectral properties, the HIDs in Figure~\ref{fig:combined_hid} show a close similarity in shape and range during the three \ixpe observations.
Moreover, the simultaneous \ixpe and \swift pointings of the Off-dip state during the third and the second observations gave consistent best-fit parameters.

\subsection{Comparison with previous \ixpe observations and combined analysis}

\begin{figure}[t]
\centering\includegraphics[width=0.8\linewidth]{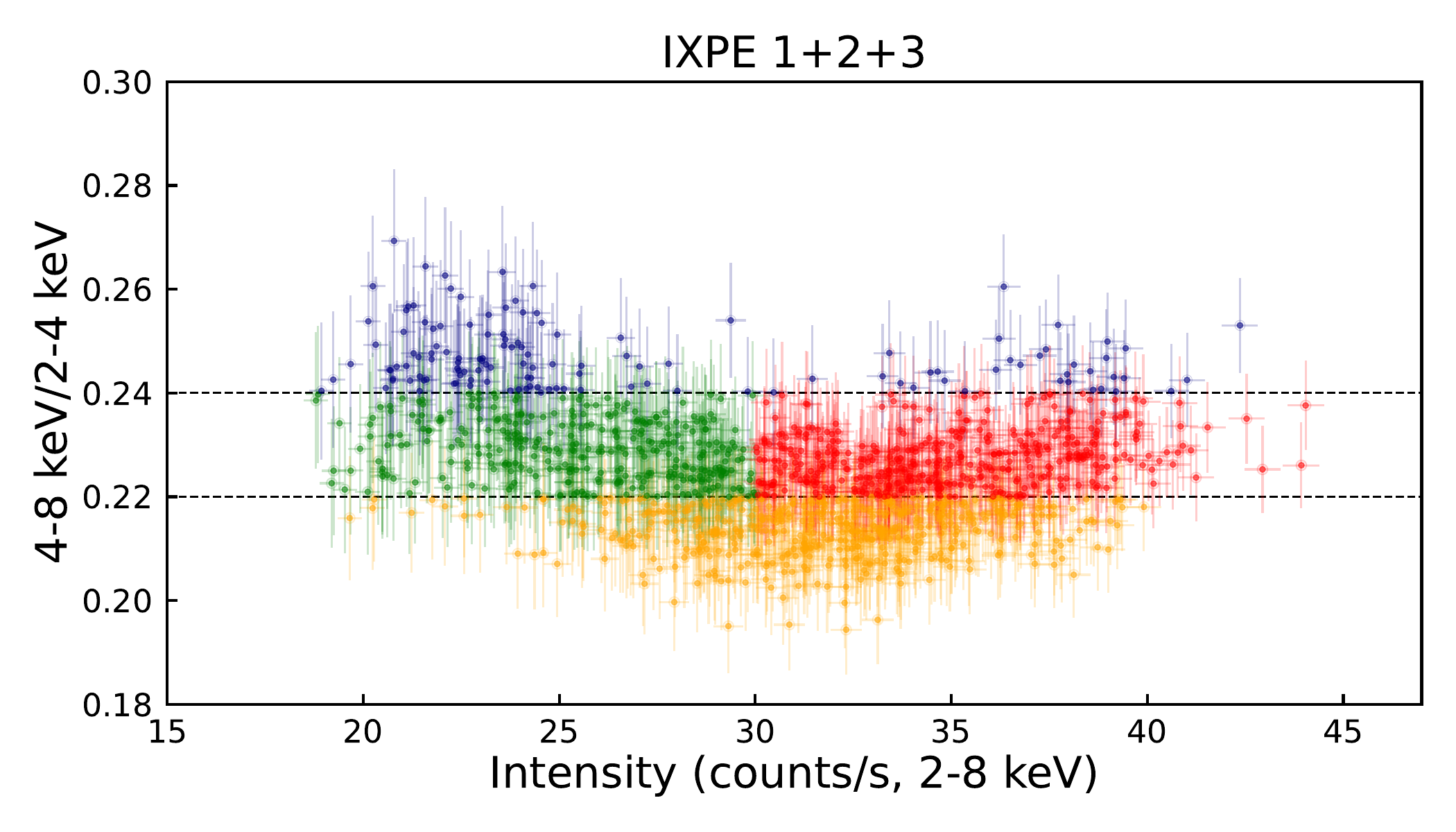}\\
    \includegraphics[width=0.8\linewidth]{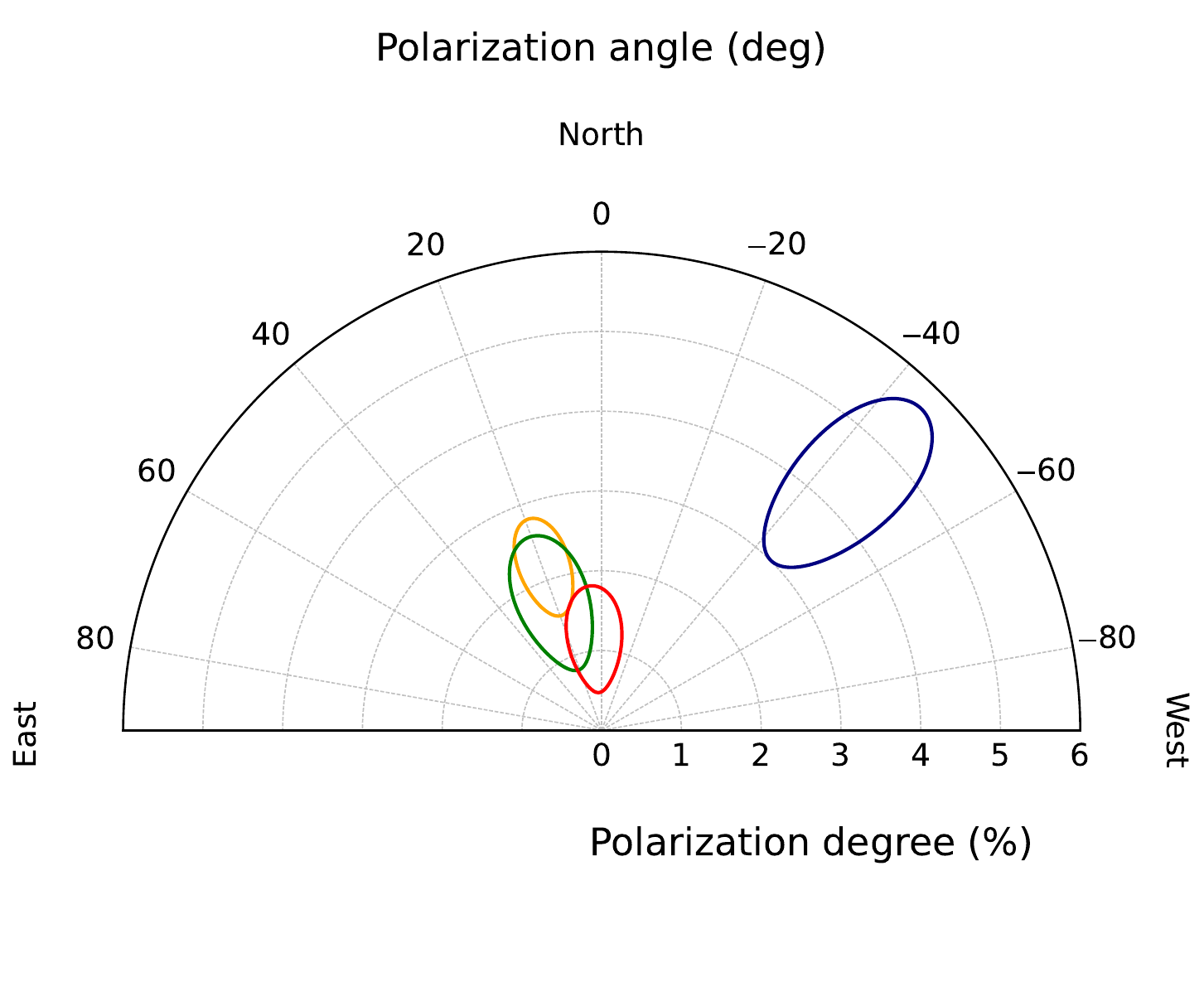}\\
    \vspace{-0.7cm}
    \caption{\source HID from the three different \ixpe observations where the different colors correspond to different HR and flux bins (top). The bottom panel reports the 90\% CL polarization contours corresponding to the four different HR bins.}
    \label{fig:hid_contour_HR}
\end{figure}

In consideration of the above similarities, we decided to combine the three \ixpe observations in a single data set in order to search with higher sensitivity for correlated variability of PD and PA with intensity and HR. We first binned the HR data, in a manner similar to \cite{Rankin2024}, and then selected the three HR intervals, almost equally-spaced, shown in Figure~\ref{fig:hid_contour_HR}.

In the higher HR interval, the one corresponding to the dips, we measured PD${\sim} 4.3\%$ at PA${\sim}-47\degr$, whereas the lower HR bin gave PD${\sim}2.4\%$ at PA${\sim}22\degr$. Thus, the total rotation was PA${\sim}70\degr$, a value similar to the continuous range spanned during the first \ixpe observation of \source. The bin with intermediate HR displayed a lower PD (${\sim}1.3\%$) and a PA compatible with that of the low HR bin. We also split this intermediate HR bin into a low and a high intensity bin (see Figure~\ref{fig:hid_contour_HR}), but no significant differences in their polarization properties were found.

\begin{figure}[t]
\centering
\includegraphics[width=0.8\linewidth]{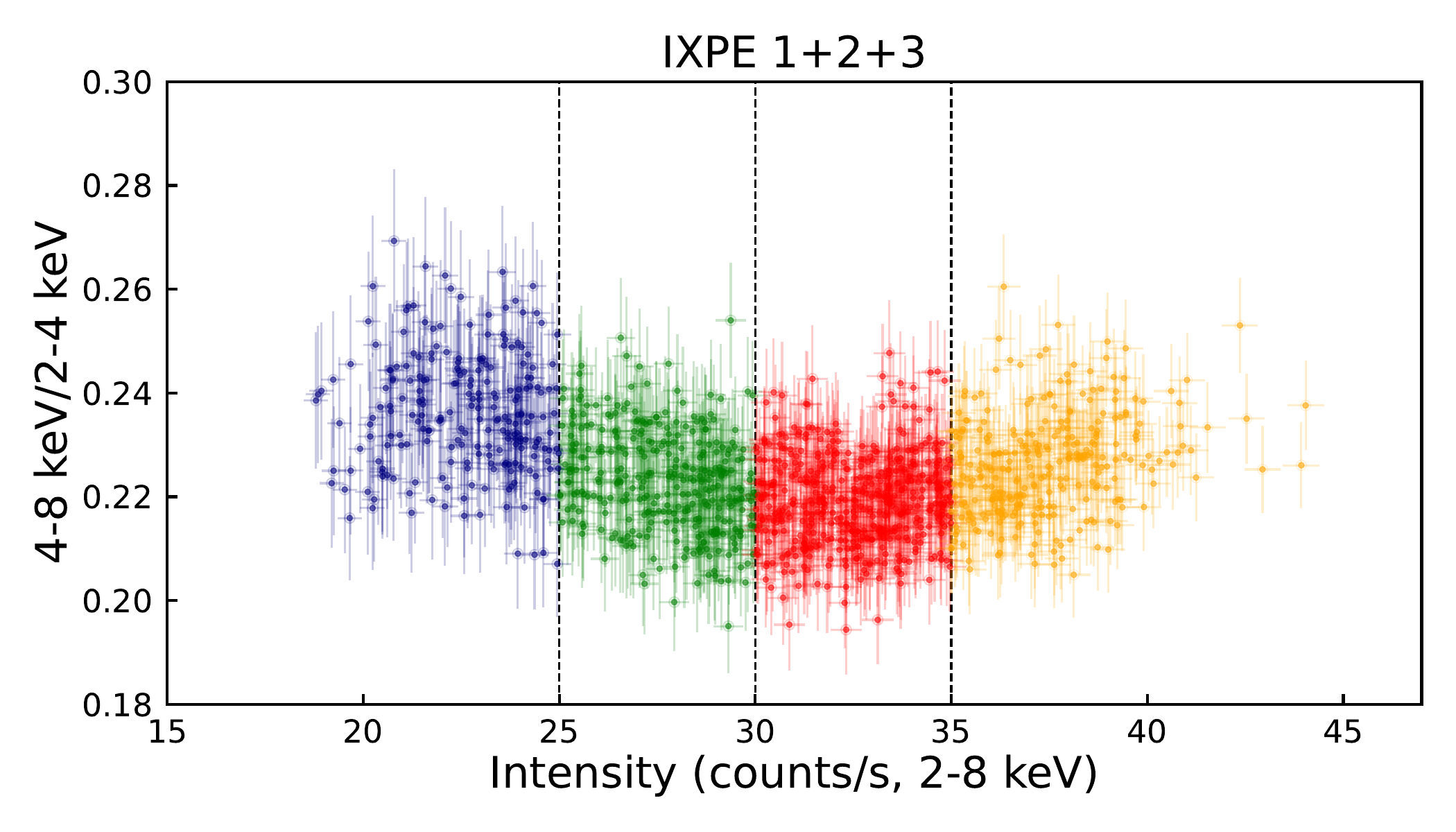} \\
\includegraphics[width=0.8\linewidth]{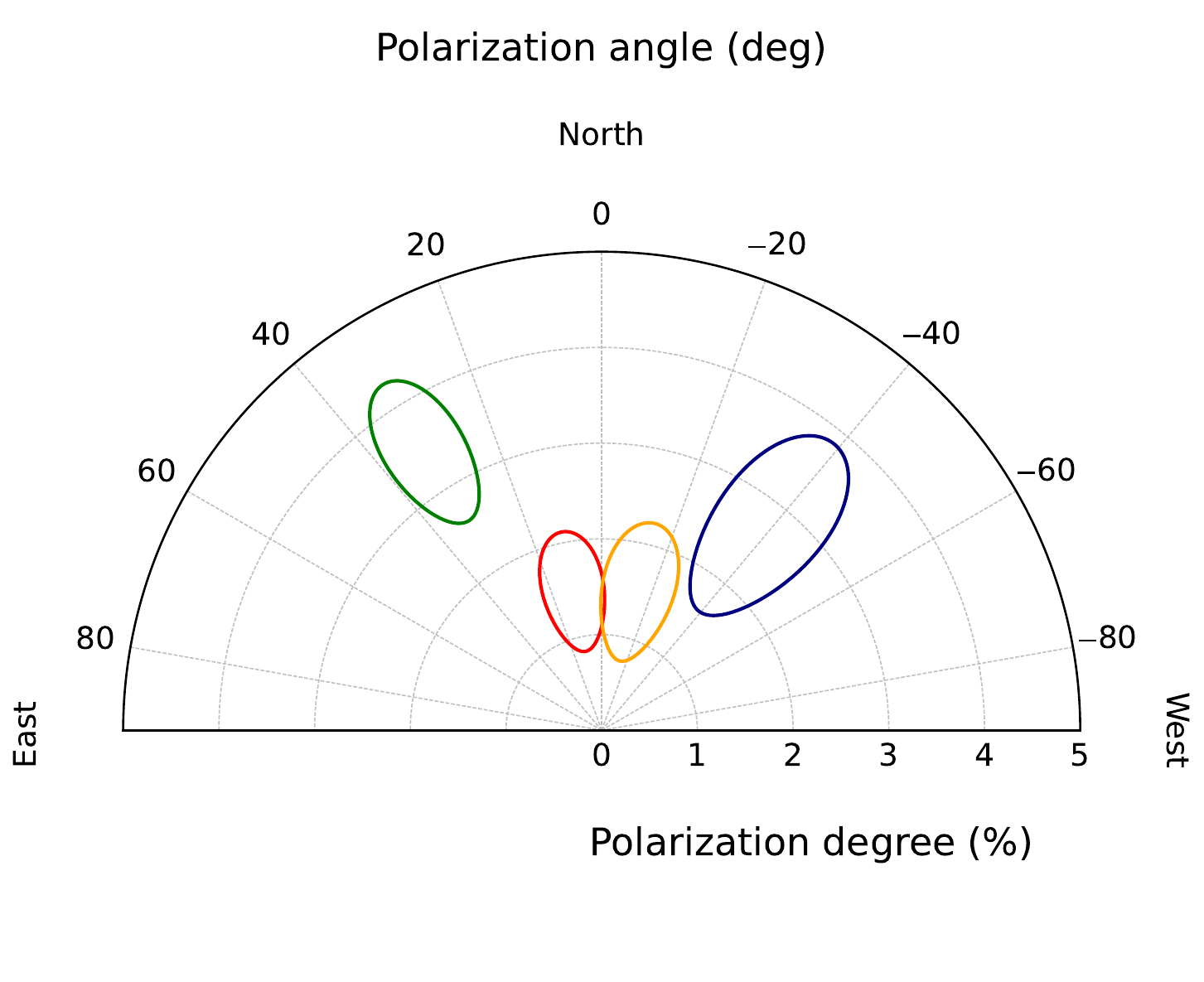}\\
\vspace{-0.7cm}
\caption{\source HID from the three different \ixpe observations where the different colors correspond to different bins in intensity (top). The bottom panel reports the polarization contours at 90\% CL, corresponding to each intensity bin.}
\label{fig:hid_contour_flux}
\end{figure}

A similar analysis was performed by dividing the combined data into four, almost equally-spaced, intensity bins, shown with different colors in Figure~\ref{fig:hid_contour_flux}. In the lowest intensity bin, we found PD${\sim}3\%$ at PA${\sim}-40\degr$. For bins of progressively higher intensities, the polarization evolved as follows: PD ${\sim}3.5\%$ at PA${\sim}32\degr$, PD ${\sim}1.5\%$ at PA${\sim}11\degr$ and finally PD ${\sim}1.5\%$ at ${\sim}-14\degr$ in the highest intensity bin. Moreover, also in this case, the total variation of PA relative to the value in the dips was ${\sim}70\degr$.

Given the size of the uncertainty contours plotted in Figure~\ref{fig:hid_contour_flux}, it can be concluded that these results are consistent with those reported in the HR binned analysis. 
In summary, low and low-intermediate intensity and high hardness were associated to a higher polarization degree; higher intensity and lower hardness corresponded to a lower polarization degree. The overall polarization angle swing across the extremes was ${\sim}70\degr$. 

\subsection{Polarization and spectral components} 

These results may be interpreted as arising from variations of the continuum spectral components. We note that the harder Comptonized component (modeled with \texttt{bbodyrad}), which is believed to originate in the BL or the SL at the NS surface, contributed about half of the total flux in the Off-dip state and about $80$\% during the dips. In fact, the softer component modeled with \texttt{diskbb} and associated with disk emission showed a more pronounced decrease in intensity during dips, leading to spectral hardening. In the \nicer spectra we analyzed, the hardening appears to arise from the combination of a lower \texttt{diskbb} temperature and increased absorption (see Table~\ref{tab:spec_pol}).\footnote{Such a spectral variation was also observed in the \xmm data and could not be readily interpreted, see \cite{Trigo12}.}

The polarization vectors of these two components, in systems with aligned BL/SL and disk symmetry axes, are expected to be ${\sim}90\degr$ apart or parallel, depending on the geometry and optical depth of the boundary/spreading layer represented by the Comptonized spectrum \citep{st85,Loktev22,Bobrikova24b}. The mixing of these two components would generally produce a different polarization vector. In this case, variations of the two spectral components, giving rise to the intensity and hardness variations, would also translate into PD and PA variations, from the sum of the two polarization vectors. In fact, for different HR values, the relative contribution of the two components changes and so does the resulting polarization vector. Intensity variations due to dips, which are associated with HR variations, can produce the same effect. This trend is visible, for example, in Figure~5 of \cite{Bobrikova24b} where energy resolved polarization shows a trend consistent with the spectral decomposition in terms of \texttt{diskbb} and \texttt{bbodyrad}: the polarization at higher energy was almost aligned with the hard component and in the softer energy bin with the polarization vector of the softer component.

Thus, during the dips the lower contribution of the soft component produces a hardening in the spectrum, and it is expected that the polarization will be dominated by the hard component. In our spectropolarimetric analysis, we found PD at ${\sim}$5\%  during the dip of the third observation. This value is higher than the maximum polarization degree of 3--4\% expected for a source inclination of 60--80\degr. For the soft \texttt{diskbb} component, a surprisingly high PD of 10\% was obtained in the dip, which should be considered with care, whereas the value out of the dip was ${<}4\%$. 
The latter limit is consistent with the maximum value of polarization expected for the disk component, which is 4\% in the \ixpe energy band \citep[{e.g.},][]{Loktev22}. Higher PD values are expected if a further polarized component is introduced in the model, such as a wind \citep[see, {e.g.},][]{Tomaru2024,Nitindala24}, or an ADC.

\subsection{Dips and polarization from ADC or disk wind}

The shallow orbital modulation \citep[see Figure~\ref{fig:maxi_folded} in this letter and Figure~16 in][]{Iaria2014} together with the presence of dips testifies that \source is a high inclination LMXB with an extended ADC and/or a disk wind, which scatters some photons from the X-ray emitting regions along our line of sight. The shape of the folded light curve holds a clear resemblance to that of other LMXBs \citep[see, {e.g.},][]{Parmar88}; its few percent amplitude together with the luminosity close to the Eddington one demonstrates that the central X-ray source remains visible at all orbital phases. The periodic dips may coincide with superior conjunction of the NS, if material lost by the donor in its vicinity occults a fraction of the X-ray radiation. The dips revealed by \ixpe occurred at phases about $-0.2$ and $0.35$ away from the periodic dips. Other LMXBs have displayed off-phase dips; for instance two-dips per orbital cycle were observed occasionally in \mbox{4U 1916$-$053} \citep{Smale1988}. Dips separated by $\sim$0.5 in phase are believed to occur when clumps form close to the point where the accretion stream from the companion impacts the rim of the disk and about 180\degr\ away from that.

The characteristic size of the clumps, $R_{\rm clump}$, can be estimated from the dip duration $T_{\rm dip}$ through the relation
\begin{equation}
    R_{\rm clump} = \frac{T_{\rm dip}}{P_{\rm orb}}R_{\rm out},
\end{equation}
where $P_{\rm orb}$ is the orbital period and $R_{\rm out} \sim (1.0-2.6)\times10^{12}$~cm is the disk outer radius \citep{dai2014}. The clumps giving rise to the two dips of the third \ixpe observation are estimated to be $\sim (5-15)$ and  $(2-6)\times 10^9$~cm in size; that is about two orders of magnitude smaller than the thickness of the disk, assuming that its half-height is 0.1--0.25 times $R_{\rm out}$. For the clumps to obscure the central X-ray source,  the line of sight to \source must almost graze the outer disk. 
An ADC size of $R_{\rm ADC}{\sim} 10^{10}$ cm was estimated in \mbox{4U 2129$+$470} \citep{McClintock82} and \mbox{4U 1822$-$37} \citep{White1982}. \source, being a more luminous source, probably drives and photoionizes a larger corona: therefore, we assume $R_{\rm ADC}{\sim} 10^{10}-10^{11}$ cm for it.
Such a radius is about ${\sim}10^3$ times larger than the regions where the bulk of the X-ray luminosity is produced. The photon source feeding the ADC can thus be considered point-like, a condition under which the scattered radiation can possess a high PD, provided that the ADC is not spherically symmetric.
%For instance, a constant density, prolate (ellipticity of ${\sim}0.3$), axially symmetric ADC fed by unpolarized photons would give rise to PD= 80\%$\times \tau$ 
%($\tau$ is the Thomson optical depth), when observed from a 60\degr\ inclination, see, {e.g.}, Figure~3 in \cite{Brown1977}, where an ellipticity ${\sim}0.3$ corresponds to $\gamma{\sim}0.6$.
For instance, a constant density, oblate (ellipticity of ${\sim}0.3$), axially symmetric ADC fed by unpolarized photons would give rise to PD= 20\%$\times \tau$
($\tau$ is the Thomson optical depth), when observed from a 60\degr\ inclination, see, {e.g.}, Figure~4 in \cite{Brown1977}, where an ellipticity ${\sim}0.3$ corresponds to
$\gamma{\sim}0.24$\footnote{$\gamma$ is given by Equation 30 in \cite{Brown1977} for $a=3$.}.

\begin{figure}%[!b]
\centering
\includegraphics[width=\linewidth]{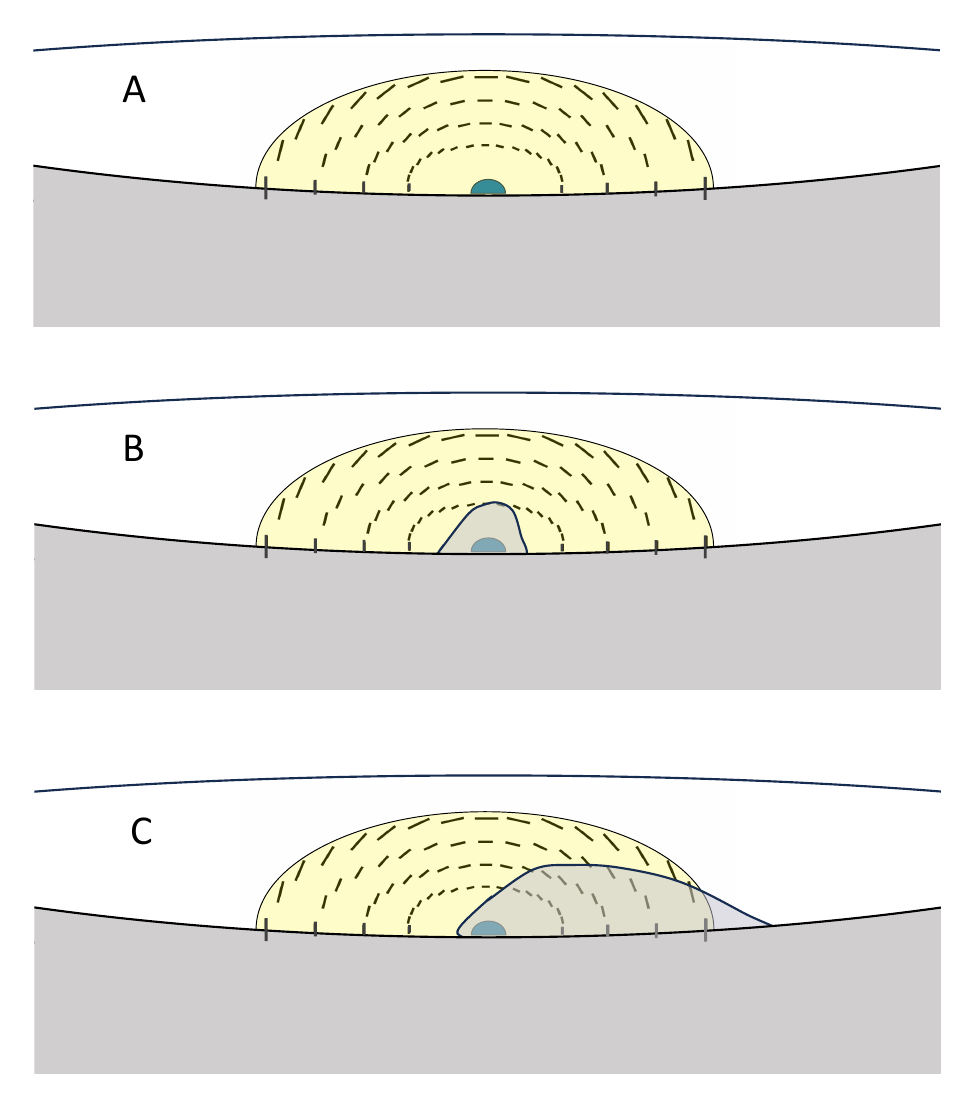}
\caption{Sketch of the proposed geometry of \source (not to scale). The gray region represents the outer disk rim, the yellow region the ADC, and the blue circle the central X-ray source. The black segments show the polarization angle of the radiation scattered by the ADC.  The polarization angle at the observer (given by the average of the polarization angles) changes from the off-dip state (panel A) to the dip state when a small clump obscures the central source and a negligible part of the ADC (panel B) and a large clump obscures the central source a fairly large fraction of the ADC (panel C). 
}
\label{fig:sketch_adc}
\end{figure}

In the case of \source (and likewise other LMXBs), the PD of the central X-ray source may thus be enhanced by scattering in an ADC, if its optical depth is, say, of order $\sim 0.1$. Besides increasing the PD to the observed level, the ADC may also cause a rotation of PA, together with an associated variation of the PD, when a clump giving rise to a dip passes in front of and obscures the central X-ray emitting region, while leaving the vision of the ADC virtually unaltered. In fact, by reducing the ratio of the (unscattered) photons reaching us directly from the central regions to those scattered along our line of sight by the ADC, the passage of the clumps will affect the mixing of the polarization vectors from the central region and the ADC leading to changes in the polarization angle.\footnote{Except for the case in which the PAs of the central source and the ADC are coaligned.} Some contribution to the rotation of the PA may also be produced when a large clump covers a significant fraction of the ADC, breaking the axial symmetry of the corona. These effects are illustrated in the sketch in Figure~\ref{fig:sketch_adc}, panels B and C.

Similar considerations would hold if the scattering region was the disk wind, rather than the non-expanding corona we outlined above \citep[see also][]{Nitindala24}. The warm absorber in the \nicer spectral fits of \source indicates column density of $\sim (5-6) \times 10^{23}$~cm$^{-2}$ in the \texttt{cabs} component corresponding to the Thomson optical depth of $\tau \sim 0.3-0.4$. A fan-shaped, 30 degree aperture wind launched from a radius of $10^{10}-10^{11}$~cm, as inferred in \cite{Trigo12}, is similar in size and height to the ADC discussed above and may give rise to comparably large PDs.\footnote{However it must be accounted for a factor $f <1$ resulting from the lower solid angle subtended by the wind at the central source; $f\sim 0.5$ for this geometry).} Most of the optical depth of a wind with density scaling $\propto r^{-3}$ is reached within a few times the launching radius, such that the size of this equivalent ``wind corona'' of \source exceeds the size of the clumps, and PA rotation can be produced in the same way as in the ADC case.

\section{Conclusions}\label{sec:conclusion}

Based on the April 2024 \ixpe observation of \source, together with coordinated observations from \nicer and \swift, we studied the evolution of the spectral and polarization properties across the off-dip and dip states. Variations in PD and PA were detected that were correlated with the source intensity and hardness variations associated with the dips. The harder Comptonized component, likely arising from the boundary/spreading layer, and the softer blackbody-like emission from the disk had
comparable fluxes in the off-dip intervals, when the PD was about 1--2\%. During the two dips, absorption by both neutral and warm gas increased, the Comptonized component became dominant and the PD raised up to 4--5\%, displaying a $\sim 70$\degr\ swing in PA. 
Joint analysis of the three \ixpe observations led to a more precise polarimetric study, which provided confirmation of the relationship of the polarization with the source intensity and hardness and of the rotation of the PA.  

Interpreting the high PD and the $\sim70$\degr\ PA variation, in light of the polarization properties expected for disk and boundary/spreading layer emission proved difficult \citep[see also][]{Bobrikova24a,Bobrikova24b}. In fact, for systems whose components (the BL/SL and the disk) have symmetry axes aligned, polarization vectors are expected to be either parallel or orthogonal, with a polarization degree lower than 4\% for each of them \citep{Loktev22,st85}. In this framework explaining PA rotation would require that the symmetry axis of the two components is offset and changing, for instance through a misalignment between the disk and the NS axis, as suggested by \citet{Bobrikova24a}.

We propose that scattering in the ADC or in the disk wind, in association with obscuration by clumps causing the dips, can drive polarization angle variations, as well as polarization degree variations, in \source as well as other LMXBs seen from high inclination. Detailed modeling would be required for this scenario, which also accounts for the energy distribution and polarization properties of the central emitting regions. 
%\begin{acknowledgments}

\section*{Acknowledgments} 

The Imaging X-ray Polarimetry Explorer (IXPE) is a joint US and Italian mission.  
The Italian contribution is supported by the Italian Space Agency (Agenzia Spaziale Italiana, ASI) through contract ASI-OHBI-2022-13-I.0, agreements ASI-INAF-2022-19-HH.0 and ASI-INFN-2017.13-H0, and its Space Science Data Center (SSDC) with agreements ASI-INAF-2022-14-HH.0 and ASI-INFN 2021-43-HH.0, and by the Istituto Nazionale di Astrofisica (INAF) and the Istituto Nazionale di Fisica Nucleare (INFN) in Italy. 
This research used data products provided by the IXPE Team (MSFC, SSDC, INAF, and INFN) and distributed with additional software tools by the High-Energy Astrophysics Science Archive Research Center (HEASARC), at NASA Goddard Space Flight Center (GSFC). 

We thank \textit{Swift} Project Scientists for approving our DDT request to observe \source. 
The authors acknowledge the \nicer team for the simultaneous observations. 
ADM and FLM are partially supported by MAECI with grant CN24GR08 “GRBAXP: Guangxi-Rome Bilateral Agreement for X-ray Polarimetry in Astrophysics”.
This research has been supported by the Finnish Cultural Foundation grant 00240328 (AB) and the Academy of Finland grant 333112 (AB, JP).

%\end{acknowledgments}

\facilities{\ixpe, \swift, \nicer}

\software{astropy \citep{2013A&A...558A..33A,2018AJ....156..123A},  
          SaoDS9, HEASoft \citep{heasoft}, Stingray 2.0.0rc2 \citep{stingray1,stingray2}
          }

\bibliography{biblio}{}
\bibliographystyle{aasjournal}

\end{document}